\let\includefigures=\iftrue

\input harvmac

\newcount\yearltd\yearltd=\year\advance\yearltd by 0

\noblackbox

\includefigures
\message{If you do not have epsf.tex (to include figures),}
\message{change the option at the top of the tex file.}
\input epsf
\def\figin{\epsfcheck\figin}\def\figins{\epsfcheck\figins}
\def\epsfcheck{\ifx\epsfbox\UnDeFiNeD
\message{(NO epsf.tex, FIGURES WILL BE IGNORED)}
\gdef\figin##1{\vskip2in}\gdef\figins##1{\hskip.5in}
\else\message{(FIGURES WILL BE INCLUDED)}%
\gdef\figin##1{##1}\gdef\figins##1{##1}\fi}
\def\DefWarn#1{}
\def\figinsert{\goodbreak\midinsert}
\def\ifig#1#2#3{\DefWarn#1\xdef#1{Fig.~\the\figno}
\writedef{#1\leftbracket Fig.\noexpand~\the\figno}%
\figinsert\figin{\centerline{#3}}\medskip\centerline{\vbox{
\baselineskip12pt\advance\hsize by -1truein
\noindent\footnotefont{\bf Fig.~\the\figno:} #2}}
\bigskip\endinsert\global\advance\figno by1}
\else
\def\ifig#1#2#3{\xdef#1{Fig.~\the\figno}
\writedef{#1\leftbracket Fig.\noexpand~\the\figno}%
\global\advance\figno by1}
\fi

\def\doublefig#1#2#3#4{\DefWarn#1\xdef#1{Fig.~\the\figno}
\writedef{#1\leftbracket Fig.\noexpand~\the\figno}%
\figinsert\figin{\centerline{#3\hskip1.0cm#4}}\medskip\centerline{\vbox{
\baselineskip12pt\advance\hsize by -1truein
\noindent\footnotefont{\bf Fig.~\the\figno:} #2}}
\bigskip\endinsert\global\advance\figno by1}

\def\triplefig#1#2#3#4#5{\DefWarn#1\xdef#1{Fig.~\the\figno}
\writedef{#1\leftbracket Fig.\noexpand~\the\figno}%
\figinsert\figin{\centerline{#3\hskip1.0cm#4\hskip1.0cm#5}}\medskip\centerline{\vbox{
\baselineskip12pt\advance\hsize by -1truein
\noindent\footnotefont{\bf Fig.~\the\figno:} #2}}
\bigskip\endinsert\global\advance\figno by1}

\noblackbox
\def\IZ{\relax\ifmmode\mathchoice
{\hbox{\cmss Z\kern-.4em Z}}{\hbox{\cmss Z\kern-.4em Z}}
{\lower.9pt\hbox{\cmsss Z\kern-.4em Z}}
{\lower1.2pt\hbox{\cmsss Z\kern-.4em Z}}\else{\cmss Z\kern-.4em
Z}\fi}
\def\IB{\relax{\rm I\kern-.18em B}}
\def\IC{{\relax\hbox{\kern.3em{\cmss I}$\kern-.4em{\rm C}$}}}
\def\ID{\relax{\rm I\kern-.18em D}}
\def\IE{\relax{\rm I\kern-.18em E}}
\def\IF{\relax{\rm I\kern-.18em F}}
\def\IG{\relax\hbox{$\inbar\kern-.3em{\rm G}$}}
\def\IGa{\relax\hbox{${\rm I}\kern-.18em\Gamma$}}
\def\IH{\relax{\rm I\kern-.18em H}}
\def\II{\relax{\rm I\kern-.18em I}}
\def\IK{\relax{\rm I\kern-.18em K}}
\def\IP{\relax{\rm I\kern-.18em P}}

\font\cmss=cmss10 \font\cmsss=cmss10 at 7pt
\def\IR{\relax{\rm I\kern-.18em R}}

\def\frac#1#2{{#1 \over #2}}

\def\OL#1{ \kern1pt\overline{\kern-1pt#1
   \kern-1pt}\kern1pt }
\def\la{\Lambda}
\def\vp{\varphi}
\def\[{\left [}
\def\({\left (}
\def\]{\right ]}
\def\){\right )}

\Title{\vbox{\baselineskip12pt
\hbox{SU-ITP-07-03}
\hbox{UCB-PTH-07/04}
 }} {\vbox{ \vskip -5cm {\centerline{Colliding with a Crunching Bubble}
} } }

\vskip  -5mm
 \centerline{Ben Freivogel$^{1}$, Gary T. Horowitz$^2$, and Stephen Shenker$^{3}$}

\bigskip
{\sl
\centerline{$^{1}$Center for Theoretical Physics, Department of
Physics, 
University of
California, and}
\centerline{Lawrence Berkeley National Laboratory}
\centerline{Berkeley, CA 94720, USA}
\medskip
\centerline{$^{2}$Department of Physics} \centerline{
University of California, Santa Barbara, CA 93106, USA}
\medskip
 \centerline{$^{3}$Department of Physics }
\centerline{ Stanford University, Stanford, CA 94305, USA}

\bigskip
\bigskip \medskip
}


\leftskip 8mm  \rightskip 8mm \baselineskip14pt \noindent
{\tenbf \hskip 50mm Abstract}
\bigskip
In the context of eternal inflation we discuss the fate of $\Lambda =
0$ bubbles  when they collide with
$\Lambda < 0$ crunching bubbles.  When the $\Lambda=0$ bubble is
supersymmetric, it is not completely destroyed by collisions.
If the domain wall separating the bubbles has higher tension than the BPS bound, it is
 expelled from the  $\Lambda = 0$ bubble and does not alter its long time behavior.  If the 
 domain wall saturates the BPS bound, then it stays inside the $\Lambda = 0$ bubble and
 removes a finite fraction of future infinity.  In this case,
 the crunch singularity is hidden behind the horizon
 of a stable hyperbolic black hole.  

\leftskip 0mm  \rightskip 0mm
 \Date{\hskip 8mm March  2007}

\lref\MossIQ{
  I.~G.~Moss,
  ``Singularity Formation From Colliding Bubbles,''
  Phys.\ Rev.\ D {\bf 50}, 676 (1994).
}
\lref\juanclaim{J. Maldacena, private communication.}

\lref\LangloisUQ{
  D.~Langlois, K.~i.~Maeda and D.~Wands,
  ``Conservation laws for collisions of branes (or shells) in general
  relativity,''
  Phys.\ Rev.\ Lett.\  {\bf 88}, 181301 (2002)
  [arXiv:gr-qc/0111013].
}

\lref\GuthPN{
  A.~H.~Guth and E.~J.~Weinberg,
  ``Could The Universe Have Recovered From A Slow First Order Phase
  Transition?,''
  Nucl.\ Phys.\  B {\bf 212}, 321 (1983).
}

\lref\ColemanAW{
  S.~R.~Coleman and F.~De Luccia,
  ``Gravitational Effects On And Of Vacuum Decay,''
  Phys.\ Rev.\  D {\bf 21}, 3305 (1980).
}
\lref\HawkingGA{
  S.~W.~Hawking, I.~G.~Moss and J.~M.~Stewart,
  ``Bubble Collisions In The Very Early Universe,''
  Phys.\ Rev.\  D {\bf 26}, 2681 (1982).
}
\lref\BlancoPilladoHQ{
  J.~J.~Blanco-Pillado, M.~Bucher, S.~Ghassemi and F.~Glanois,
  ``When do colliding bubbles produce an expanding universe?,''
  Phys.\ Rev.\  D {\bf 69}, 103515 (2004)
  [arXiv:hep-th/0306151].
}
\lref\DouglasES{For a review see 
  M.~R.~Douglas and S.~Kachru,
  ``Flux compactification,''
  arXiv:hep-th/0610102.
}

\lref\DeWolfeUU{
  O.~DeWolfe, A.~Giryavets, S.~Kachru and W.~Taylor,
  ``Type IIA moduli stabilization,''
  JHEP {\bf 0507}, 066 (2005)
  [arXiv:hep-th/0505160].
}

\lref\censustaker{
A. Maloney, S. Shenker and L. Susskind, in preparation.}

\lref\FreivogelXU{
  B.~Freivogel, Y.~Sekino, L.~Susskind and C.~P.~Yeh,
  ``A holographic framework for eternal inflation,''
  Phys.\ Rev.\  D {\bf 74}, 086003 (2006)
  [arXiv:hep-th/0606204].
}

\lref\KhanVH{
  K.~A.~Khan and R.~Penrose,
``Scattering of two impulsive gravitational plane waves,''
  Nature {\bf 229}, 185 (1971).
}

\lref\CentrellaIB{
  J.~Centrella and R.~A.~Matzner,
  ``Colliding Gravitational Waves In Expanding Cosmologies,''
  Phys.\ Rev.\  D {\bf 25}, 930 (1982).
}

\lref\BrillMF{
  D.~R.~Brill, J.~Louko and P.~Peldan,
  ``Thermodynamics of (3+1)-dimensional black holes with toroidal or higher
  genus horizons,''
  Phys.\ Rev.\  D {\bf 56}, 3600 (1997)
  [arXiv:gr-qc/9705012].
}

\lref\VanzoGW{
  L.~Vanzo,
  ``Black holes with unusual topology,''
  Phys.\ Rev.\  D {\bf 56}, 6475 (1997)
  [arXiv:gr-qc/9705004].
}

\lref\BlauCW{For a review of thin wall techniques in General Relativity see e.g., 
  S.~K.~Blau, E.~I.~Guendelman and A.~H.~Guth,
  ``The dynamics of false vacuum bubbles,"
  Phys.\ Rev.\  D {\bf 35}, 1747 (1987).
}

\lref\CeresoleIQ{
  A.~Ceresole, G.~Dall'Agata, A.~Giryavets, R.~Kallosh and A.~Linde,
  ``Domain walls, near-BPS bubbles, and probabilities in the landscape,''
  Phys.\ Rev.\  D {\bf 74}, 086010 (2006)
  [arXiv:hep-th/0605266].
}

\lref\BeckerKS{
  K.~Becker, M.~Becker, C.~Vafa and J.~Walcher,
  ``Moduli stabilization in non-geometric backgrounds,''
  arXiv:hep-th/0611001.
}

\lref\MicuRD{
  A.~Micu, E.~Palti and G.~Tasinato,
  ``Towards Minkowski vacua in type II string compactifications,''
  arXiv:hep-th/0701173.
}

\lref\KachruAW{
  S.~Kachru, R.~Kallosh, A.~Linde and S.~P.~Trivedi,
  ``De Sitter vacua in string theory,''
  Phys.\ Rev.\  D {\bf 68}, 046005 (2003)
  [arXiv:hep-th/0301240].
}

\lref\HawkingFZ{
  S.~W.~Hawking and I.~G.~Moss,
  ``Supercooled Phase Transitions In The Very Early Universe,''
  Phys.\ Lett.\  B {\bf 110}, 35 (1982).
}

\lref\BoussoGE{
  R.~Bousso, B.~Freivogel and I.~S.~Yang,
  ``Eternal inflation: The inside story,''
  Phys.\ Rev.\  D {\bf 74}, 103516 (2006)
  [arXiv:hep-th/0606114].
}

\lref\StarobinskyFX{
  A.~A.~Starobinsky,
  ``Stochastic de Sitter (Inflationary) Stage in the Early Universe,"
{\it  In *De Vega, H.j. ( Ed.), Sanchez, N. ( Ed.): Field Theory, Quantum Gravity and Strings*, 107-126}
}

\lref\LindeSK{
  A.~D.~Linde,
  ``Hard art of the universe creation (stochastic approach to tunneling and
  baby universe formation),''
  Nucl.\ Phys.\  B {\bf 372}, 421 (1992)
  [arXiv:hep-th/9110037].
}
\lref\Goenner{
H.~Goenner, ``Einstein tensor and generalizations of Birkhoff's theorem,"
Commun. Math. Phys. {\bf 16}, 34 (1970).
}

\lref\Barnes{
A.~Barnes, ``On Birkhoff's theorem in general relativity,"
Commun. Math. Phys. {\bf 33}, 75 (1973).
}
\lref\HawkingVC{
  S.~W.~Hawking,
  ``Black holes in general relativity,''
  Commun.\ Math.\ Phys.\  {\bf 25}, 152 (1972).
}
\lref\DafermosHQ{
  M.~Dafermos and A.~D.~Rendall,
  ``Strong cosmic censorship for surface-symmetric cosmological spacetimes with
  collisionless matter,''
  arXiv:gr-qc/0701034.
}

\lref\DeWolfeNS{
  O.~DeWolfe, A.~Giryavets, S.~Kachru and W.~Taylor,
  ``Enumerating flux vacua with enhanced symmetries,''
  JHEP {\bf 0502}, 037 (2005)
  [arXiv:hep-th/0411061].
}

\lref\CveticVR{For a review of domain walls in supergravity see
  M.~Cvetic and H.~H.~Soleng,
  ``Supergravity domain walls,''
  Phys.\ Rept.\  {\bf 282}, 159 (1997)
  [arXiv:hep-th/9604090].
}
\lref\CarvalhoFC{
  C.~Carvalho and M.~Bucher,
  ``Separation distribution of vacuum bubbles in de Sitter space,''
  Phys.\ Lett.\  B {\bf 546}, 8 (2002)
  [arXiv:hep-ph/0207275].
}

\lref\BoussoXY{
  R.~Bousso,
  ``A Covariant Entropy Conjecture,''
  JHEP {\bf 9907}, 004 (1999)
  [arXiv:hep-th/9905177].
}

\newsec{Introduction}

There is now strong evidence for the existence of a ``landscape" of vacua in string theory \DouglasES .
These include states with positive, negative and zero cosmological constant $\Lambda$.
In a static situation the negative $\Lambda$ vacua are associated to  Anti de Sitter (AdS)  geometries, and the  zero $\Lambda$ ones to flat space.  Both  can be supersymmetric.  The positive
$\Lambda$ states correspond to de Sitter (dS) geometries and describe inflating universes.  

 It is natural to ask about  the ultimate fate of a universe that starts
in an initial positive  $\Lambda$ state.   This universe will inflate, but bubbles of other vacua
will nucleate within it.  Typically the nucleation rate will be slow enough so that inflation will be eternal
and some parts of the universe will continue inflating forever.   The bubbles of $\Lambda=0$
supersymmetric vacua that are nucleated seem  very stable and so will last forever.  The bubbles
of $\Lambda <0$ vacua, even if supersymmetric, are not.  In this
cosmological context they end in a time of order $(-\la)^{-1/2}$ in a
curvature singularity we will call a crunch \ColemanAW .

Bubbles do not evolve in isolation;  they collide with each other.  In fact each bubble collides
with an infinite number of other bubbles \GuthPN .  
At first sight it appears that the collision of a crunching bubble with an asymptotically flat ($\Lambda =0$)
bubble would necessarily have dramatic consequences. Once the singularity of the crunch enters the asymptotically flat region, it might seem that there are only two possibilities:  the singularity could continue across all space causing the flat  bubble to end in a crunch,  or it could end, violating cosmic censorship (see Fig. 1).
Cloaking the singularity behind a horizon seems to violate Hawking's theorem \HawkingVC\ stating that four dimensional black hole horizons must be spherical.

\ifig\question{Bubbles with zero and negative cosmological constant can collide. The heavy black line denotes  the curvature singularity which always forms in the $\la <0$ bubble. The $\la=0$ bubble always includes a piece of future null infinity. We will investigate what happens to the future of the collision.
  }{\epsfxsize4.0in\epsfbox{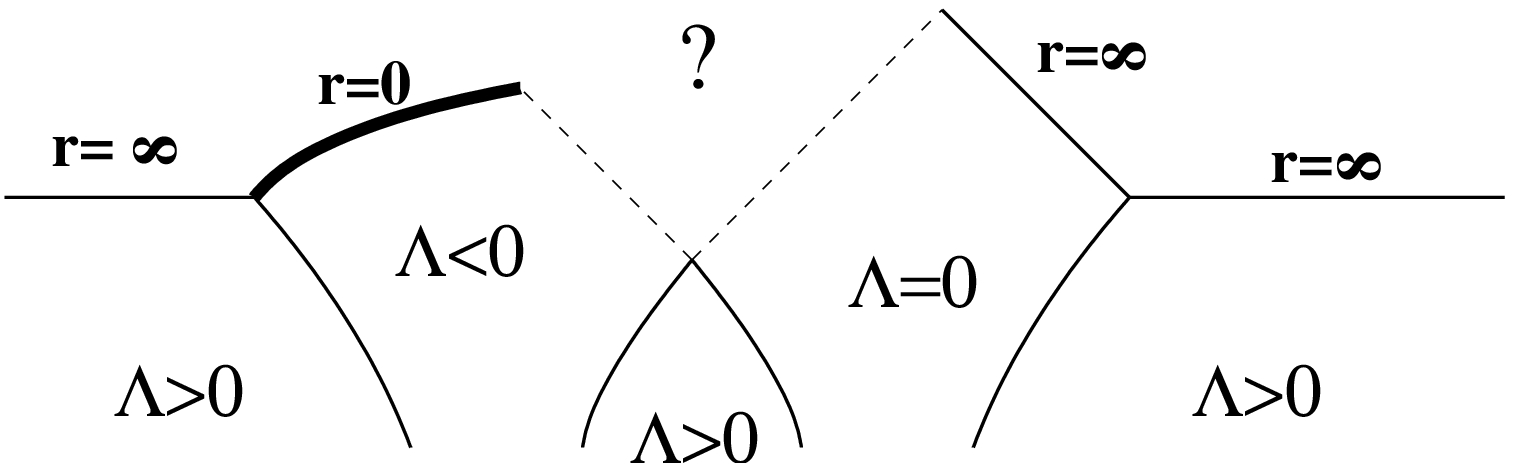}}

We study these collisions and show that while it is possible to find situations where the singularity cuts off all space,  they correspond to an unphysical choice of parameters which cannot arise in a supersymmetric theory.  When the $\Lambda =0$ bubble is supersymmetric the collision is much more mild.   This is consistent with our intuition that
supersymmetric states are stable. 

If the  tension of the domain wall \CveticVR\ connecting the crunching bubble and the flat one exceeds the 
minimum allowed by supersymmetry, the BPS bound, then  the singularity never really  enters the flat bubble. The boundary between the two bubbles accelerates away from the  flat bubble.  If the tension
saturates the BPS bound then 
the singularity does enter the flat bubble, but it remains inside a { \it hyperbolic} black hole. 
The hyperbolic symmetry of the black hole follows from the symmetry of the problem -- expanding spherical bubbles intersect in hyperbolas. Hawking's theorem is evaded since it requires that energy densities are always positive, which is not the case when the crunching bubble has $\Lambda < 0$. 

Supersymmetry is not crucial to the above argument, except to
bound the tension of the domain wall. Put another way, supersymmetry
guarantees that the Minkowski vacuum is completely stable. Even without
invoking supersymmetry, we find that if the static
Minkowski vacuum is stable, bubbles of the Minkowski
vacuum will be stable cosmologically, in the sense that they are not
destroyed by crunching bubbles. In our analysis 
large tension in the
final domain wall guarantees that the wall expands out to infinite size.
This may hold true more generally; it would be interesting to ask if the 
results of \BlancoPilladoHQ, who consider colliding domain walls in a 
slightly different context, would change for sufficiently large tension.

In the next section we analyze the bubble collisions using a thin wall
approximation in an effective field theory of gravity coupled to a
scalar field. This includes a description of the appropriate
background solutions with hyperbolic symmetry and matching conditions.
(Some details are relagated to an appendix.) 
In section 3 we estimate various parameters that appear in the problem and discuss when the thin wall approximation is valid. In
section 4 we consider possible complications such as going beyond the thin wall approximation and violating the hyperbolic symmetry. Finally, we discuss the effect of multiple collisions in section 5.

\newsec{Effective Field Theory Description and Thin Wall Approximation}

\subsec{General framework}

To begin our investigation of the collision of these bubbles, we consider gravity coupled to a scalar field with potential $V(\phi)$. We require that $V$ have three local  minima, one with $V>0$, one with $V=0$ and one with $V<0$. This will allow us to consider a solution that starts in de Sitter space and nucleates bubbles with both zero and negative cosmological constant. Since we expect our underlying theory to be supersymmetric, we will also require that V can be derived from a superpotential via
\eqn\superpot{V = 2W'^2 - 3{W^2 \over m_p^2}}
Given $V$, this is a first order ODE for $W$ which almost always has solutions. However we require that $W'=0$  at the local minimum where $V=0$. This imposes a restriction on $V$  relating the depth of the negative minimum to the potential barrier separating it from the $V=0$ minimum.

The spacetime describing a single bubble in four dimensions has $SO(3,1)$ symmetry. A solution with two bubbles has a preferred direction connecting the center of the bubbles. However it still preserves an $SO(2,1)$ symmetry acting orthogonally. This means that the colliding bubble spacetime is analogous to a spherically symmetric spacetime, with the two-dimensional spheres replaced by two-dimensional hyperboloids. In general, the hyperboloids could be timelike or spacelike, but we will see that in the region of interest, the hyperboloids are always spacelike, so our metrics will depend on only one time and one space direction. 

\subsec{Background metrics}

To study what happens after the collision, we will use a thin wall approximation. The conditions under which this is justified will be discussed in the next section. In this approximation, the scalar field is constant and sitting at a local minimum of the potential on both sides of the wall. So the spacetimes can be constructed by gluing together solutions with the appropriate cosmological constant and hyperbolic symmetry.  We now review these solutions. 

For spherical symmetry, there is a well known uniqueness theorem. It turns out that there is a similar uniqueness theorem for hyperbolic symmetry \refs{\Goenner,\Barnes}. In other words, there are no hyperbolic gravitational waves. The unique solution
with negative cosmological constant is  \refs{\VanzoGW,\BrillMF}
\eqn\neglmetrics{
ds^2 = -f(r) dt^2 + {dr^2\over f(r)} + r^2 dH_2^2 }
where 
\eqn\deff{f(r) = {r^2\over  \ell^2} -1 -{ 2GM\over  r} }
and
\eqn\dH{ dH_2^2 = d\rho^2 + \sinh^2\rho d\vp^2}
is the metric on a unit hyperboloid. When $M=0$, this metric is simply AdS with radius $\ell$ in hyperbolic coordinates, and $r=0$ is a coordinate singularity. For all $M\ne 0$, $r=0$ is a curvature singularity.  

 These solutions have the unusual property
that there is a horizon even when the mass parameter $M$  is negative,
provided $GM > -\ell/3\sqrt 3$. For all $M$ above this bound, \neglmetrics\
describes a black hole with horizon at $f=0$. This is an infinite
area, hyperbolic event horizon. It is clear that Hawking's theorem
stating 
four dimensional black hole
horizons must be spherical does not apply when the energy density can
be negative.
For $M$ more negative than the bound,
the solution has a naked timelike singularity and no horizon. 
When $M$ is negative but above the bound, the singularities are timelike and 
there is an inner horizon as well as the event horizon. 
The positive mass solutions have a spacelike singularity and no inner
horizon. When $M$ is very small,  the horizon is close to the AdS
radius, 
so all positive mass black holes have a horizon radius $r_h > \ell$.  
Even allowing for negative $M$, the horizon radius cannot be
arbitrarily small. 
All black holes have $r_h \ge \ell /\sqrt 3$. The temperature of these black holes is
\eqn\tempBH{  T = {3r_h^2-\ell^2\over 4\pi r_h \ell^2}}
which is positive for all  black holes. One can check that the specific heat is also positive,
so these black holes can be in thermal equilibrium with their Hawking
radiation for both positive and negative $M$.

In terms of the AdS/CFT correspondence, these black holes are dual to a thermal state of the CFT on a hyperboloid. At first sight this is puzzling since conformal invariance requires that the scalars have a term in their lagrangian  proportional to  $\phi^2 {\cal R}$ where ${\cal R}$  is the scalar curvature of the boundary metric. If ${\cal R} <0$ there appears to be an instability. However, modes on a hyperboloid have a mass gap, and this is just enough to compensate for the negative scalar curvature and remove the instability.  An easy way to see that this cancellation must occur is to note that the boundary metric $ds^2 = -dt^2 + dH_2^2$ is conformal to $ds^2 = -d\hat t^2 + \hat t^2 dH_2^2$ which is just flat space. Since the dual field theory is conformally invariant, there cannot be an instability.

We now turn to solutions with $\la=0$. The only vacuum solution with hyperbolic symmetry is
\eqn\zerolmetrics{
ds^2 = -{ dt^2\over h(t)} + h(t) dz^2  + t^2 dH_2^2 }
where
\eqn\defh {h(t) = 1 - {t_0\over t}}
Setting $t_0=0$ yields flat spacetime in Milne-like coordinates. For positive $t_0$ there 
 is a timelike curvature singularity at $t=0$ (see Fig. 2). Since the singularities are timelike, this spacetime does not describe a black hole. Nevertheless, we refer to this solution as the ``hyperbolic Schwarzschild" solution, because the metric resembles that of an ordinary Schwarzschild black hole.

 \ifig\flatbh{The Penrose diagram for the ``hyperbolic Schwarzschild"
 spacetime. Each point represents a hyperboloid and the singularities
 are timelike. We will only be interested in the region to the future
 of the surface $t=t_0$ (denoted by the dashed line),  so these
 singularities will not enter our discussion. The ``wedges'' will be
 explained in section 2.5.}{\epsfxsize1.5in\epsfbox{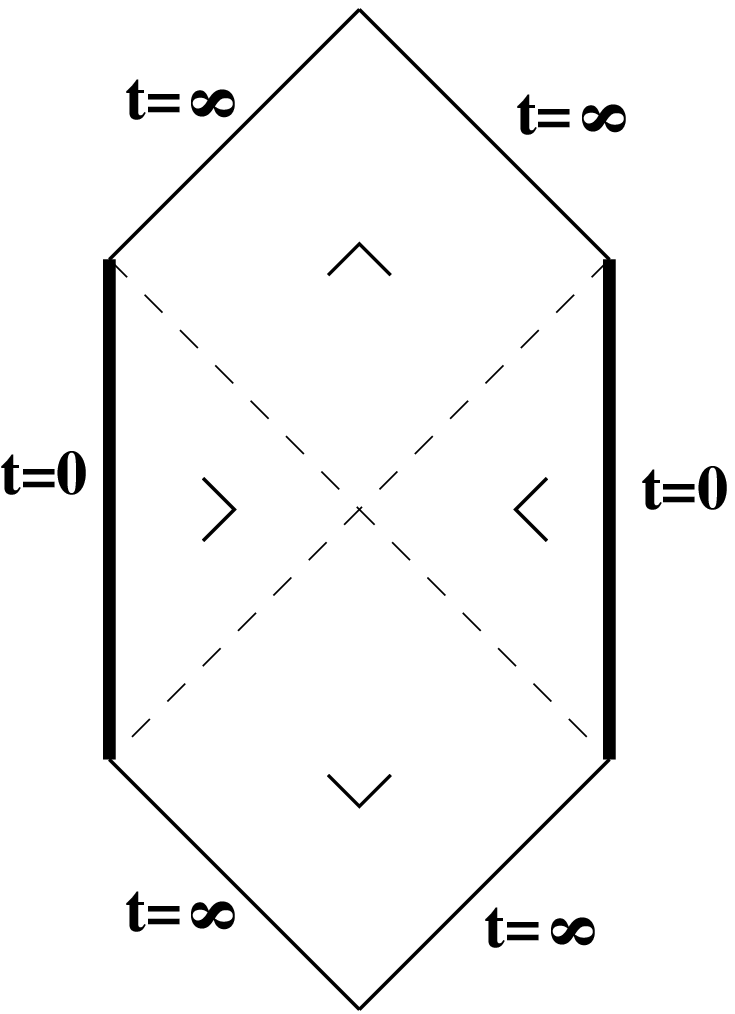}}
 
 \subsec{Matching the solutions}

We initially have two thin walls in a de Sitter background: one surrounding a bubble of Minkowski space and the other surrounding a bubble of AdS. Each wall is expanding and its intrinsic geometry is a three dimensional (timelike) hyperboloid. They collide in a two dimensional (spacelike) hyperboloid. After the collision, a new domain wall emerges separating the $\Lambda =0$ and $\Lambda < 0$ bubbles. In addition, scalar radiation can be emitted by the collision and enter each bubble. For now, we will model this as an instantaneous shell of null fluid, and consider generalizations later.  Since the $H_2$ symmetry must be preserved, the effect of this radiation is simply to allow $M$ and $t_0$ to be nonzero in the metrics after the radiation is emitted.

\ifig\pien{In our thin wall approximation, the spacetime near the collision is divided into five regions. The dotted lines denote outgoing shells of radiation. The solid lines denote two incoming bubble walls and the outgoing domain wall.}
{\epsfxsize2.0in\epsfbox{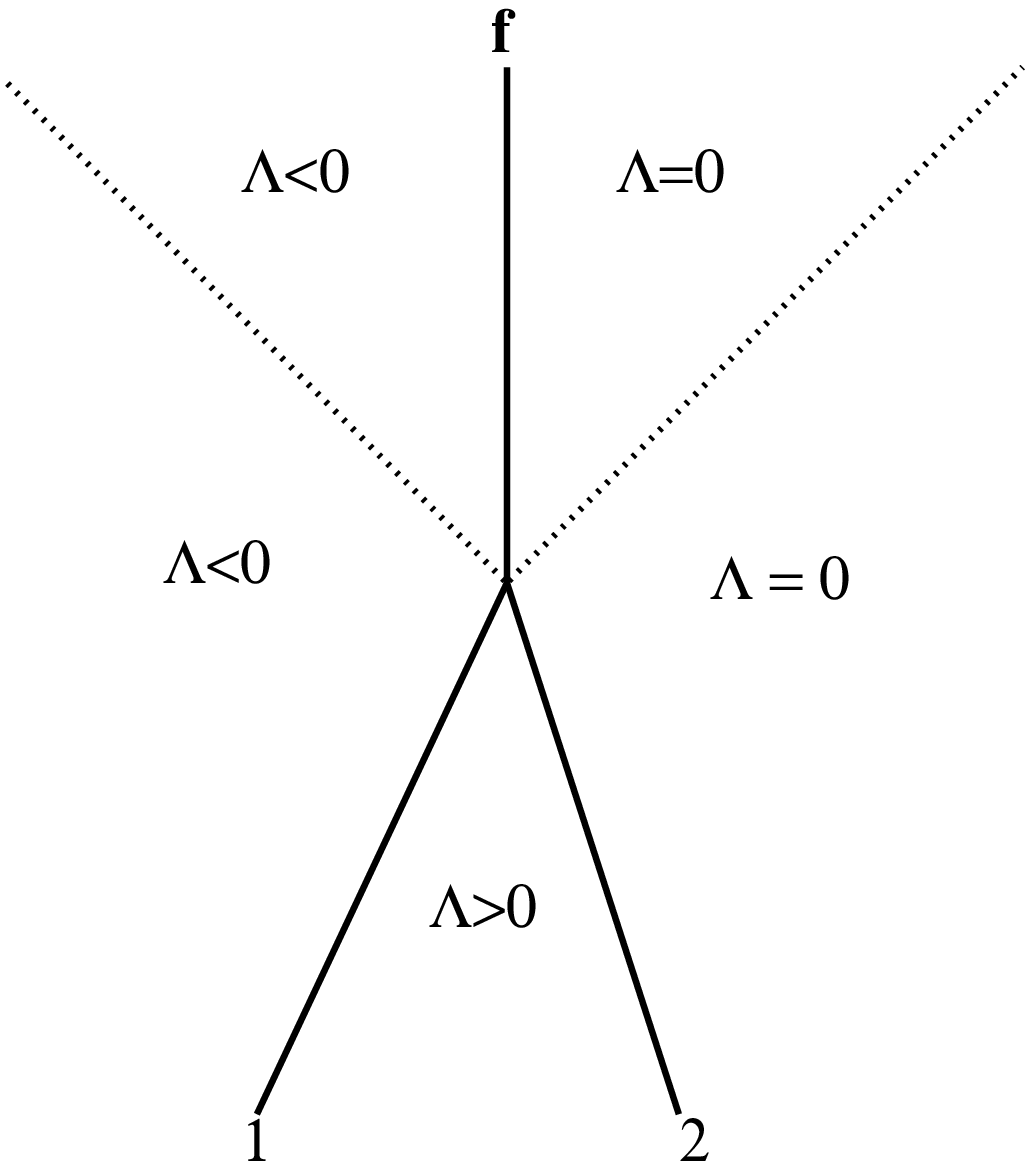}}

The general situation is shown in  \pien. We have five different regions of spacetime separated by five thin shells. The two incoming bubble walls are denoted 1 and 2, while the final wall is denoted f. We first discuss the matching across the two null shells, and then turn to the timelike ones. Let $l^\mu$ be tangent to the null geodesic generators of the shell. The stress tensor takes the form
\eqn\nullT{ T_{\mu\nu} = \sigma l_\mu l_\nu \delta(shell)}
 Since the metrics of interest have $SO(2,1)$ symmetry, there is a unique null vector $n^\mu$  orthogonal to $H_2$ normalized so that $n_\mu l^\mu = -1$. Let $h_{ij}$ be the two dimensional metric on the hyperboloid, and set $k= h^{ij} \nabla_i n_j$. The matching condition across the shell is that the metric is continuous, but $k$ is discontinuous \refs{\MossIQ,\BlauCW} 
 \eqn\nullmatch{ k_1 - k_2 = 8\pi G \sigma}

On the $\la=0$ side, we are matching \zerolmetrics\ with $t_0=0$ to  \zerolmetrics\ with nonzero $t_0$. Since $t$ is the radius of the hyperboloid, continuity of the metric means that  $t$ is continuous across the shell. In the $(t,z)$ coordinates, our null vectors have components
\eqn\nullh{ l = a (1, h^{-1}), \qquad n = {1\over 2a} ( h, -1)  }
where $a$ is a constant reflecting the freedom to boost our null basis. A short calculation yields $k =  h/a t$, so the matching condition \nullmatch\ yields
\eqn\nullansh{ {t_0\over a t^2} = 8\pi G\sigma(t)}
The fact that $\sigma$ falls off like $1/t^2$ is expected from conservation of $T_{\mu\nu}$. We can relate $t_0$ to the energy released in the bubble collision as follows. At the time of the collision, suppose an observer following the final domain wall has four-velocity
 $u^\mu$, and the hyperboloids have radius $R$.  Then if we fix $a$ by requiring $u \cdot l = -1$, $\sigma(R)$ denotes the energy density emitted into the $\la=0$ bubble, and $t_0 =  8\pi a R^2 G\sigma(R)$.
 
The matching on the $\la <0$ side is similar. Now we are matching \neglmetrics\ with $M=0$ to \neglmetrics\ with $M>0$.  Continuity of the metric requires that $r$ be continuous across the shell. In $(t,r)$ coordinates, the null vectors have components
\eqn\nullf{ l = b(f^{-1}, -1) \qquad n={1\over 2 b} (1, f)  }
where $b$ is a constant. It turns out that $k = f/br$, so the matching condition yields
\eqn\nullansf{ {2M\over b r^2} = 8\pi \sigma(r)}
If we fix $b$ so that $u \cdot l = -1$ as above, $M = 4\pi bR^2\sigma(R)$.

We now turn to the matching across the timelike thin wall that remains after the bubbles collide.  This determines the dynamics of this domain wall and controls the outcome of the collision in this approximation. (Note that we also need to impose conservation of energy-momentum
at the collision. We discuss this in the appendix.)
 We are now matching \neglmetrics\ to \zerolmetrics\ along a timelike
 surface. We will denote the radii of the hyperboloids as a function
 of proper time on the domain wall by $R(\tau)$ so the metric on the
 wall is simply $ ds^2 = -d\tau^2 + R(\tau)^2 dH_2^2$. One of the
 matching conditions is that the metric is continuous. This already
 has an important consequence. 
Any timelike surface in the region $t>t_0$ of \zerolmetrics\
 must have 
$t$ monotonically increasing, and hence $R(\tau)$ must also be
 monotonically increasing. 
We will show in section 2.5 that the
 collision $always$ occurs in the region $t>t_0$.
This means that the thin wall in \neglmetrics\ must always  expand
 outward. It cannot fall into the black hole.
 
 The second thin wall matching condition is that the extrinsic curvature is discontinuous:
\eqn\matching{\Delta k_{ij} = -8\pi G (S_{ij} - {1\over 2} g_{ij} S)}
where $S_{ij}$ is the stress tensor on the wall and $ij$ run only over the three dimensions tangent to the wall. Assuming $S_{ij}$ takes the form of a perfect fluid, the components of $k_{ij}$ in an orthonormal basis satisfy
\eqn\matchcomp{ \Delta k_{00} = - 4\pi G (\rho + 2p),  \qquad \Delta k_{11} = \Delta k_{22} = - 4\pi G\rho}
In addition, conservation of the stress energy tensor implies 
\eqn\conser{\dot \rho = -2(\rho + p) {\dot R\over R}}
We will start by considering the case that the stress tensor is dominated by a cosmological constant, so $\rho + p = 0 $ and $\rho$ is constant.

From the $\la <0$ side, we have 
\eqn\kf{ k_{11} = {1\over r} ( \dot r^2 + f)^{1/2}  }
while from the $\la =0$ side we get 
\eqn\kh{ k_{11} = \pm {1\over t} ( \dot t^2 - h)^{1/2}  }
where the upper sign applies to walls moving in the $+z$ direction and the lower sign is for walls moving in the $-z$ direction.
On the domain wall, $r(\tau) = t(\tau) = R(\tau)$, so the second junction condition in \matchcomp\ is 
\eqn\finalcond{( \dot R^2 + f)^{1/2} \mp  ( \dot R^2 - h)^{1/2} = \kappa R}
where we have set $\kappa = 4\pi G\rho$.  Squaring this twice yields
\eqn\poteq{\dot R^2 + V_{eff}(R) =0}
where
\eqn\pot{V_{eff}(R) = -h(R) - {[f(R)+h(R)-\kappa^2 R^2]^2 \over 4\kappa^2 R^2 }}
If this equation is satisfied, the other junction condition in \matchcomp\ is automatically satisfied as well. Plugging in the definitions of $f$ and $h$ from \deff\ and \defh, the effective potential is
\eqn\explpot{V_{eff}(R) = -1 + {t_0\over R} - {\[(\kappa^2 - \ell^{-2})R^2 + (2GM + t_0) R^{-1}\]^2\over 4\kappa^2 R^2}}

We are mainly interested in the behavior for large $R$. This divides into three cases, depending on whether $\kappa < 1/\ell, \ \kappa > 1/\ell$, or $\kappa = 1/\ell$. In  the first two cases, $V_{eff}(R) \approx -\lambda^2 R^2$ for some constant $\lambda$, so $R(\tau) = e^{\lambda \tau}$. The domain wall  accelerates out to large radius. 
On the $\la<0$ side, the domain wall actually makes it out to infinity in finite $t$ (Fig. 4). 
\ifig\adsbh{If $\kappa \ne 1/\ell$, the domain wall accelerates out to infinity in finite time in the AdS-Schwarzschild black hole. As we explain later, we expect a singularity to form along the heavy dashed line.}{\epsfxsize2.0in\epsfbox{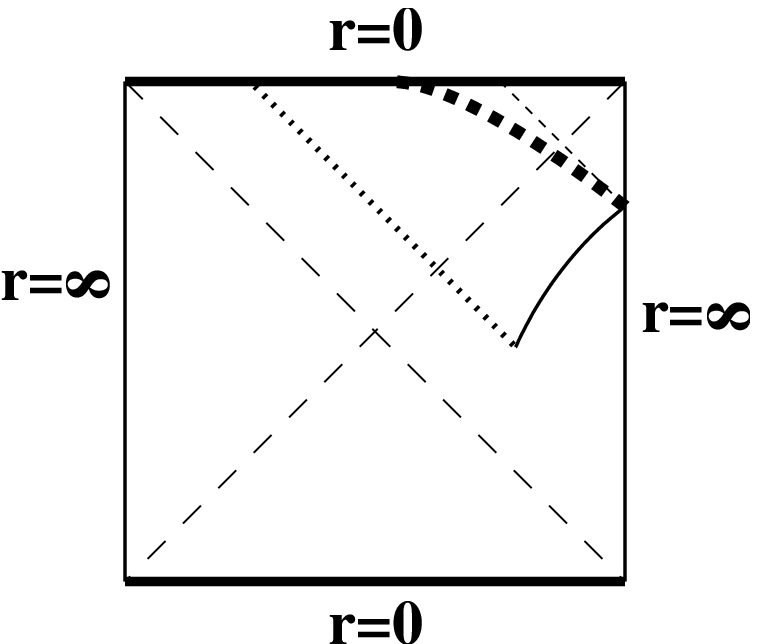}}
\noindent This is because
a unit timelike vector $(\dot t,\dot r)$ in  \neglmetrics\ satisfies 
\eqn\finitet{\dot t = \({1\over f }+ {\dot r^2\over f^2}\)^{1/2}}
so with $r(\tau) = R(\tau) = e^{\lambda \tau}$, the total change in $t$ is
\eqn\deltat{\Delta t \propto \int_{t_1}^\infty  {d\tau\over r(\tau)} < \infty}

On the $\la=0$ side, the domain wall accelerates out to null infinity. With $H_2$ slicing, there are two null infinities, one corresponding to $z\rightarrow \infty$ and the other with $z\rightarrow - \infty$ (\flatbh).  If we assume the collision takes place at some $z<0$, then a $\kappa < 1/\ell$ wall accelerates to $z=+\infty$, cutting off the entire $\la=0$ bubble, while a $\kappa > 1/\ell$ wall accelerates to $z=-\infty$, leaving most of the $\la=0$ bubble untouched. This follows from the matching condition \finalcond. Since we expect the region on the $\la<0$ side to end in a big crunch, the first case results in a disastrous collision. Fortunately, this case is not possible in supersymmetric theories as $\kappa = 1/\ell$ is the BPS bound. In fact, if $\kappa < 1/\ell$, then flat space can nucleate bubbles of $\la<0$ and decay, and there are solutions with arbitrarily negative energy. The more physical situation is the second: $\kappa > 1/\ell$. The complete  spacetime in this case is shown in Fig. 5 and the pieces which go into it are shown in Fig. 6. Since the domain wall accelerates away from both bubbles, the collision is very mild. The singularity never really enters the asymptotically flat bubble. 

\ifig\summary{A sketch of the conformal diagram for a collision in
which the resulting domain wall has tension exceeding the BPS
bound. In the thin wall approximation, region (a) is a piece of AdS
spacetime, (b) is a piece of the hyperbolic AdS black hole, (c) is a piece of the
hyperbolic Schwarzschild solution, while (d)
is a piece of Minkowski spacetime. Going infinitesimally beyond the
thin wall approximation results in additional singularities to create
a diagram like the one shown here. In the next figure, we show
 the regions (a), (b), and (c) in more detail.}
{\epsfxsize4.0in\epsfbox{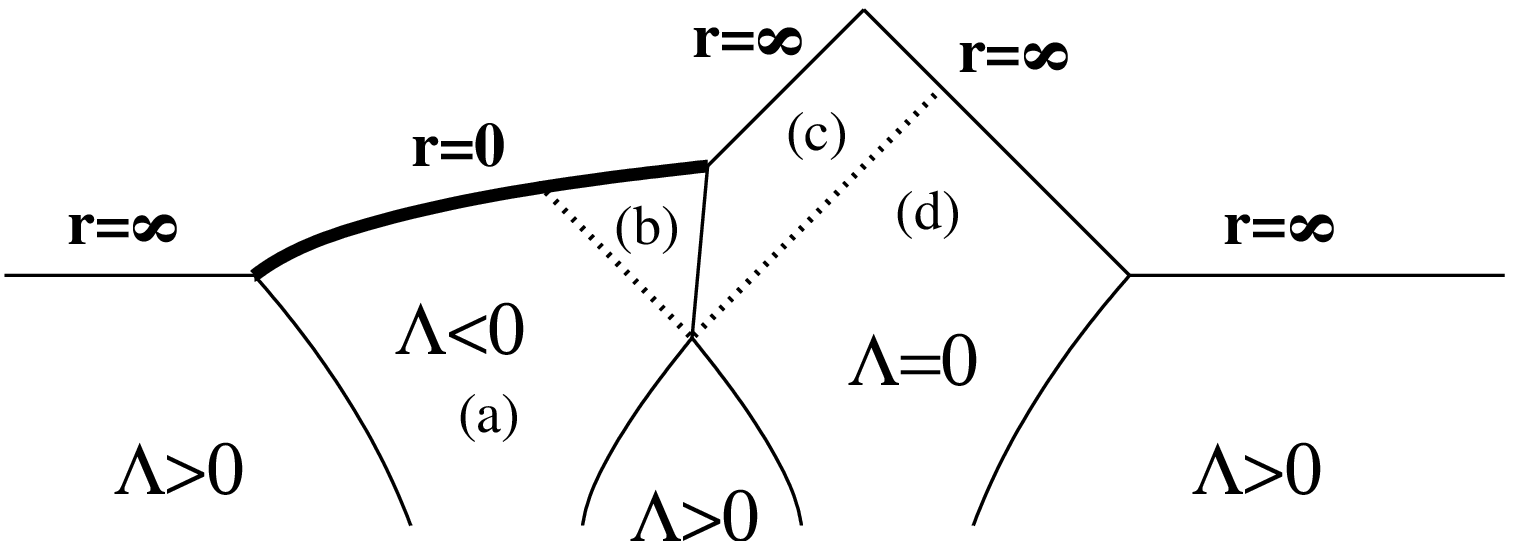}}
\triplefig\bigr{The full diagram, \summary, is constructed
from the pieces shown here. In the thin wall approximation, (a) is the
conformal diagram for AdS
spacetime in $H_2$ slicing. The top and bottom dot-dashed lines are
coordinate singularities where the size of the $H_2$ goes to zero.
 The region we use, labelled ``keep'' in the
figure,
 is bounded by the domain
wall (heavy curved lines) and the shell of radiation emitted from the
collision (dotted line). As described in the text, the thin wall
solution has a Cauchy horizon (thin, short dashes). Going beyond the
thin wall approximation, the Cauchy horizon will be removed by a
spacelike curvature singularity (heavy dashed line). Diagram (b) is
the hyperbolic AdS black hole. The region we use is bounded by the
null shell of radiation and the domain wall. The curvature singularity
of (a) must continue across the null shell into (b). Just as in (a), a
Cauchy horizon forms at the point where the domain wall hits the
boundary, so again a curvature singularity must cut off the boundary
at the point where the domain wall hits it. Diagram (c) is the
hyperbolic $\Lambda=0$ solution, bounded by the domain wall and the 
null shell radiated into the $\Lambda=0$ side of the domain wall. 
Note that we do not use the part of the spacetime
which contains timelike singularities.
}
{\epsfxsize2.0in\epsfbox{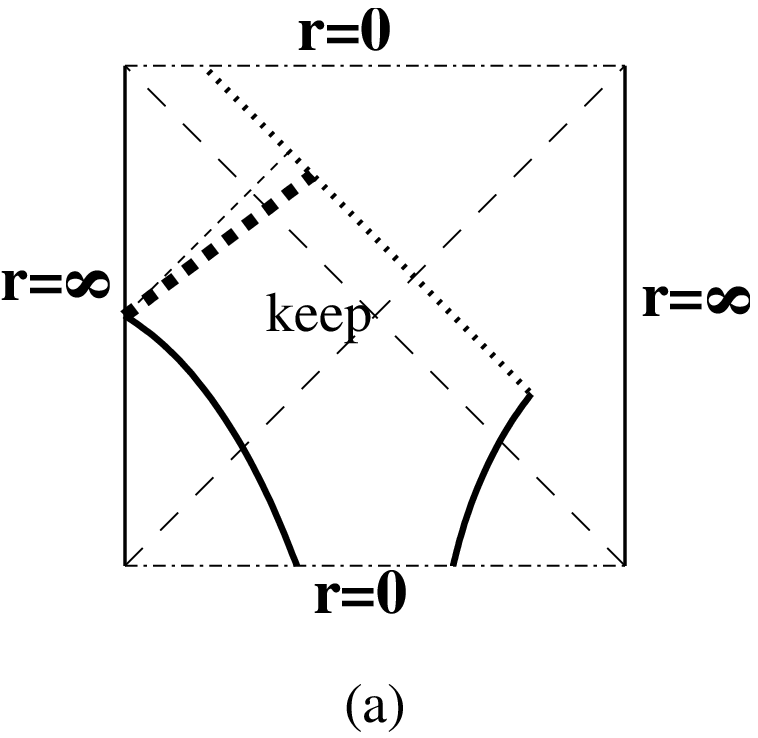}}
{\epsfxsize2.0in\epsfbox{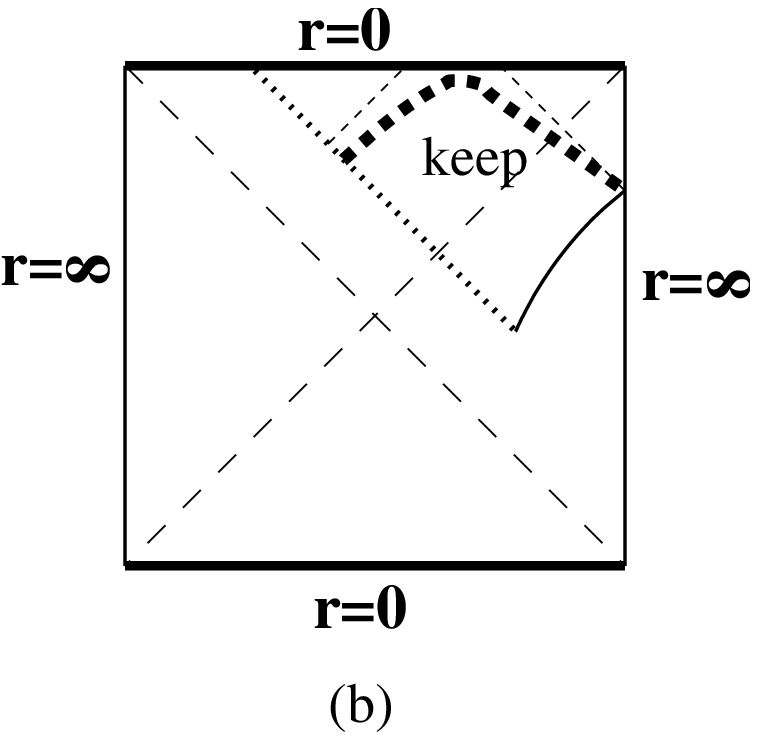}}
{\epsfxsize1.5in\epsfbox{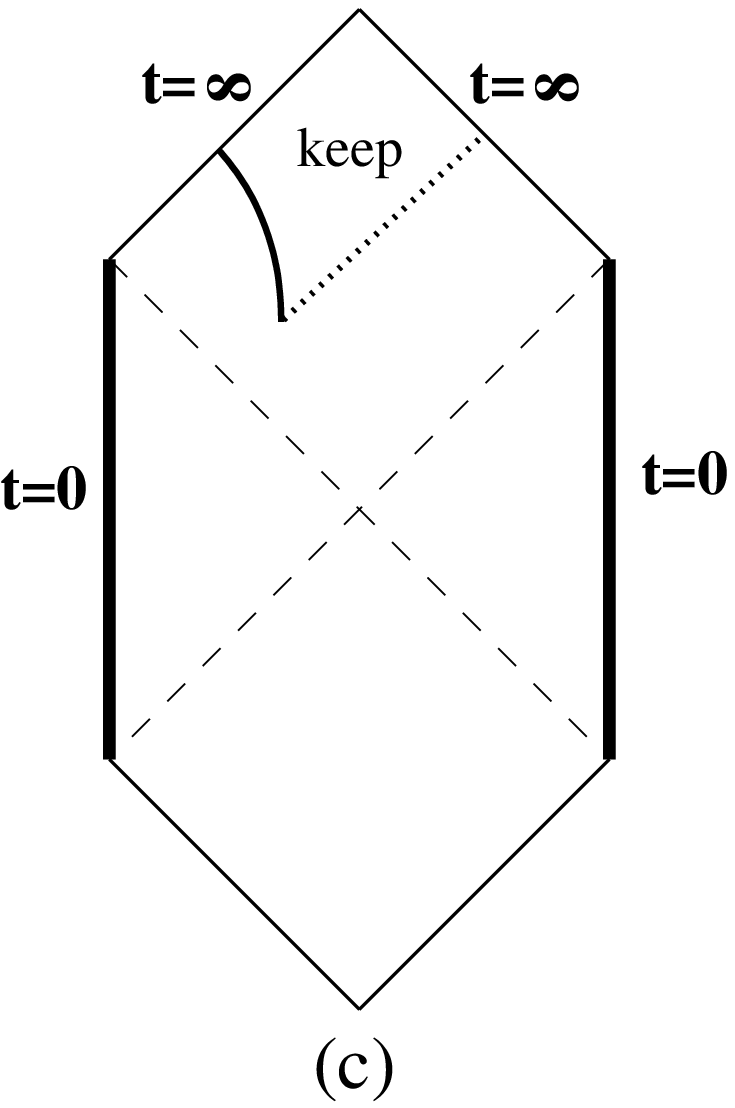}}

The third case, when the tension saturates the BPS bound ($\kappa = 1/\ell$), is qualitatively different. Now, $V_{eff} = -1$ at large $R$, so $R(\tau)=\tau$. On the $\la<0$ side, the domain wall expands much more slowly and reaches infinite $t$ (since \deltat\ is now logarithmically divergent) as shown in  Fig. 7.  On the $\la=0$ side, the domain wall now follows a constant $z$ surface at late time and reaches timelike infinity. The induced metric on the wall is now flat.  Since the effect of nonzero $t_0$ and $M$ are both negligible at large $R$, this solution approaches a BPS domain wall where flat space is glued to AdS along a constant radial surface in Poincare coordinates. 
 With nonzero $t_0$ and $M$ included,  gluing these spacetimes together yields an asymptotically flat spacetime with a hyperbolic black hole (Fig. 8). 
  
 \ifig\bpsadsbh{If the tension saturates the BPS bound, the domain wall  reaches infinity in infinite time outside the AdS-Schwarzschild black hole.}{\epsfxsize1.5in\epsfbox{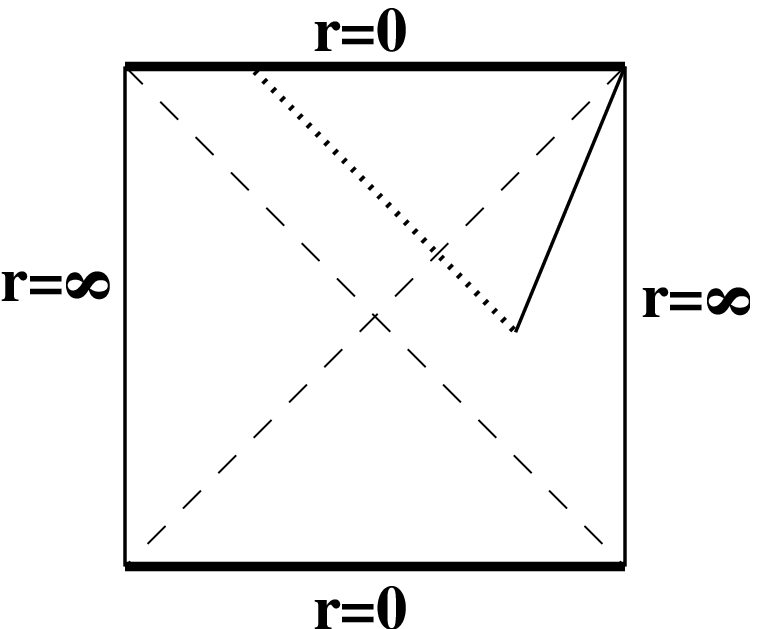}} 
 
 \ifig\bps{Gluing  \bpsadsbh\ onto the asymptotically flat bubble yields a hyberbolic black hole  with an asymptotically flat region of spacetime.} {\epsfxsize4.0in\epsfbox{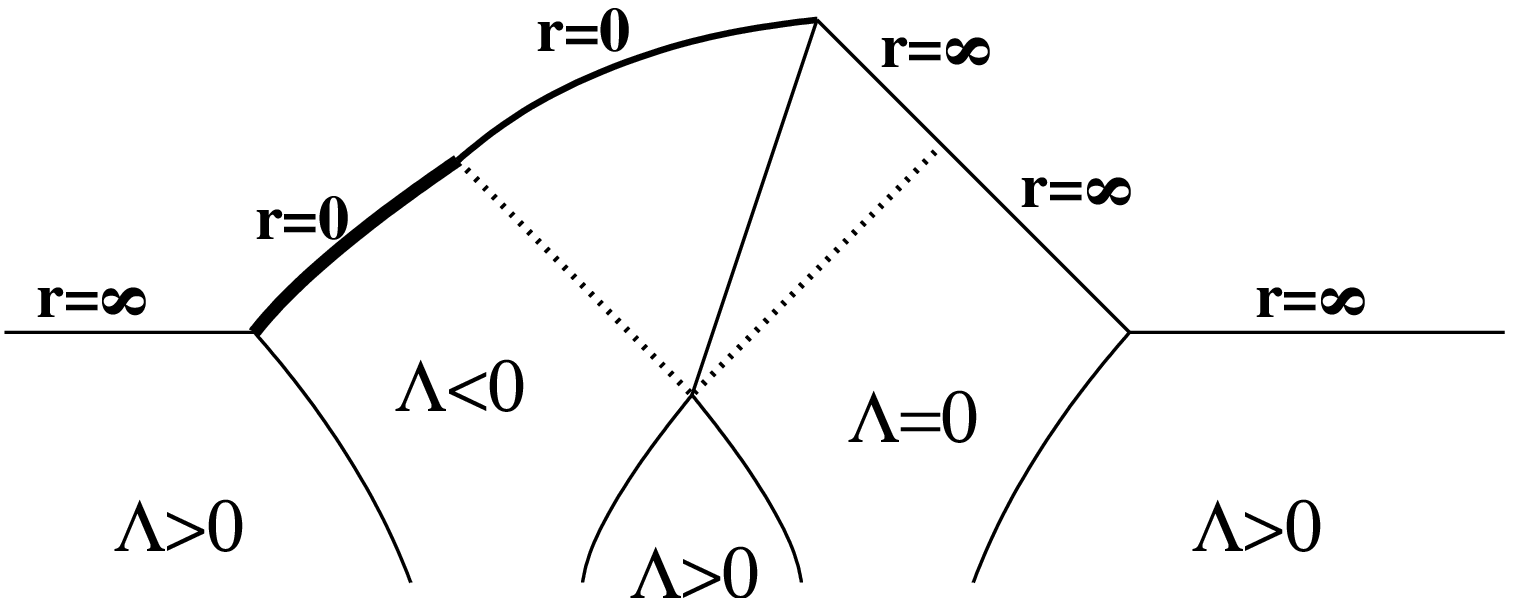}}

In the strict thin wall limit, there is no crunch, since the scalar field is assumed to lie at the local minimum of the potential. We now ask where  the crunch will be if we relax this assumption. In the $\kappa > 1/\ell$ case, we have seen that the domain wall hits infinity in finite $t$. Since the AdS infinity is timelike, there is a Cauchy horizon. Boundary conditions would have to be specified on this boundary in order for evolution to proceed past this point. Cauchy horizons are usually unstable. One version of cosmic censorship says that generically they turn into singularities. In our case, even a small amount of scalar radiation leaking off the domain wall at late time will blue-shift and build up on the Cauchy horizon causing it to become singular.  This is analogous to the instability of the inner horizon of Reissner-Nordstrom. Even if there is no classical radiation, quantum acceleration radiation will be emited by the accelerating domain wall, and this can cause the Cauchy horizon to become singular. So even though the domain wall is far from the black hole singularity, we expect a singularity to reach infinity at the same point as the domain wall (see \adsbh.).

We still have to consider the possibility that the crunch might hit the domain wall before it reaches the boundary at infinity. Fortunately, it is easy to show that this cannot happen.  Suppose it did. The point where the singularity hits the domain wall is determined from data inside its past light cone. If we consider a spacelike surface soon after the null shell is emitted, the evolution is determined by the scalar field on a finite extent of $r$ outside the horizon of the black hole. But we know that a scalar field outside the black hole decays. It does not blow up in finite time. 

\triplefig\smallr{In the case that the collision occurs on an $H_2$
with radius of curvature less than the AdS scale, the situation is
slightly different from \bigr. In this case, the null shell of
radiation on the $\Lambda<0$ side of the domain wall ends up behind the
white hole horizon. Also (b) is now a hyperbolic AdS black hole with
negative $M$.}
{\epsfxsize2.0in\epsfbox{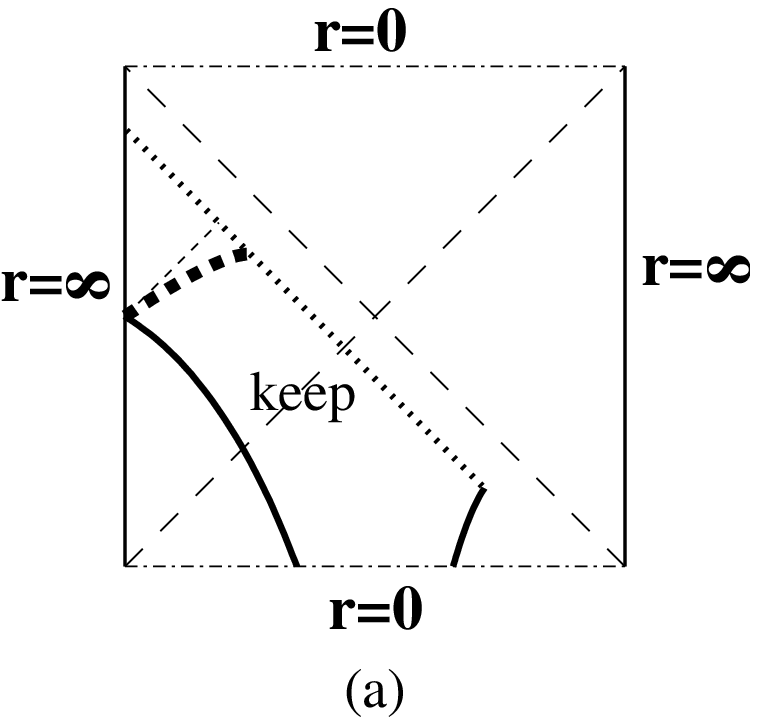}}
{\epsfxsize2.0in\epsfbox{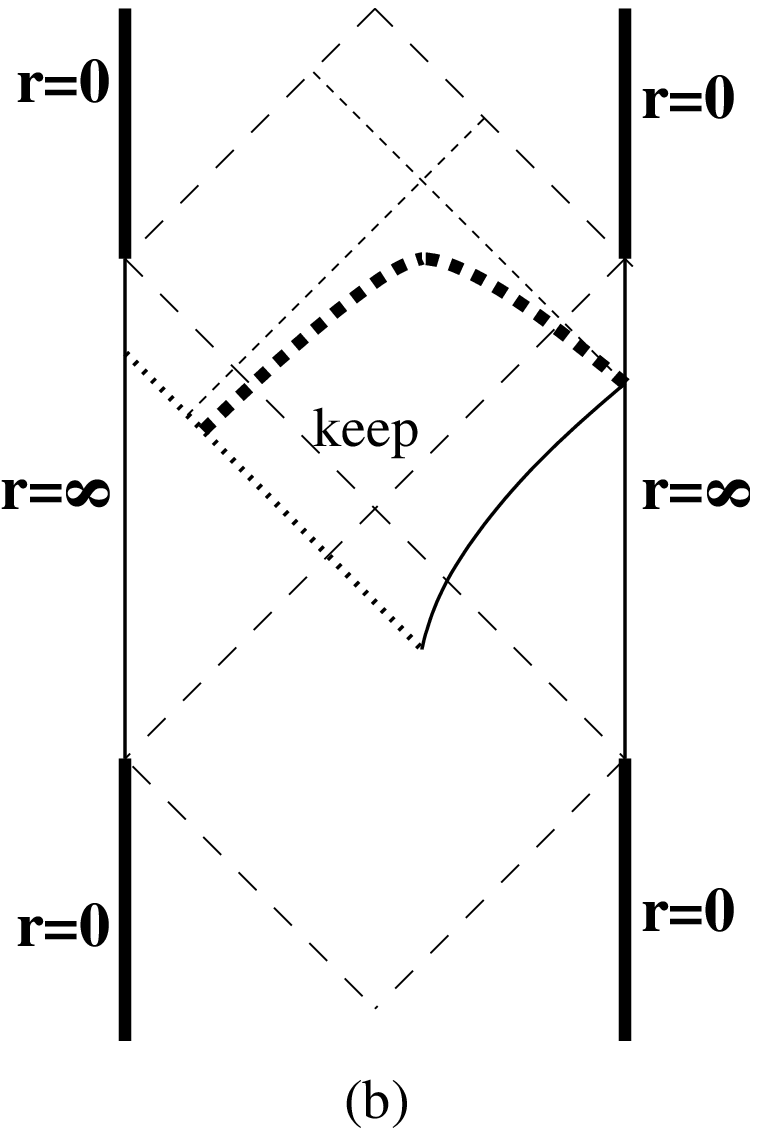}}
{\epsfxsize1.5in\epsfbox{flatbhbigr.eps}}
In the discussion so far, we have assumed  that the collision takes
place outside the horizon of the $\la <0$ black hole. However, we will
see in section 2.5 that this is not always the case. It is possible
for the collision to occur inside the past ``white hole" horizon. (We
will see that it cannot occur inside the future horizon.) In this case, the
constant $b$ in \nullf\ must be negative in order for $l,n$ to remain
future directed null vectors. This implies that $M < 0$, even though
the energy density $\sigma (R)$  radiated from the collision is
positive. In other words, some collisions produce a negative mass
black hole!
  However the black holes with $M < 0$ are not really exotic: As we
  mentioned earlier, from the outside, they behave just like the
  positive mass ones as long as $M$ is not too negative. If the tension is above the BPS bound, the spacetime is again
described by \summary, 
where the pieces are now given in Fig. 9. If the tension equals the
BPS bound, 
the spacetime will again form a hyperbolic black hole and look like \bps.

\subsec{Hyperbolic black holes}

We have seen that a bubble collision can produce a new type of asymptotically flat black hole with a hyperbolic horizon. This occurs when the domain wall tension saturates the BPS bound.  In this case, the big crunch singularity joins onto the black hole singularity inside the horizon. (The precise way that this happens would require a numerical relativity calculation.) One novel feature of this solution is that we evade the conventional wisdom that bubbles of negative cosmological constant always crunch. Since the domain wall moves out to larger and larger radii, future timelike infinity of the AdS space survives.

 Even though the spacetime is asymptotically flat far from the horizon, this black hole will  not Hawking evaporate completely. This is because the spacetime near the horizon resembles the hyperbolic AdS black hole. That black hole has positive specific heat and after evaporating a small amount of energy, it comes into thermal equilibrium with its Hawking radiation. The domain wall separating this AdS black hole from the asymptotically flat spacetime is moving out. Once the domain wall is far enough so that radiation at the black hole temperature is reflected off the AdS curvature, it cannot escape even if the spacetime is modified at larger radii.
 
 We can estimate the rate of energy loss to the flat space region using the dual CFT picture.  If the  domain wall is at radial coordinate $r\gg \ell$ this corresponds to field theoretic momentum scales $\sim r$.  The probability of 
having quanta with such momenta in the thermal black hole state is $\sim \exp(-r/T)$.  Energy loss
 to the flat space region is controlled by this factor.   The domain wall location depends on field theory time.
 For the BPS case we have $r(t) \sim \exp(t)$ and so energy loss is $\sim \exp(-\exp(t)/T)$.  This integrates to a finite total energy loss and so the black hole is stable.  Similar considerations show that
various nonperturbative effects are suppressed  even more strongly.

\subsec{Location of the collision}
We have shown in the previous sections that if the domain wall gets 
to large $R$, its late time
behavior will be determined by its tension relative to the BPS bound. 
Naively, it seems that it
should be possible to prevent the domain wall from getting to large
$R$ at all. For example, one could try to make the energy of the
collision very large so that the domain wall ends up behind a black
hole horizon and must shrink down to zero size.
Indeed, if the energy density of the radiation emitted
onto the $\Lambda=0$ side of the domain wall is
large enough, then the domain wall will be behind the horizon of the
hyperbolic Schwarzschild solution - a confusing region with a timelike
singularity. 

Luckily, it turns out that the simple assumption that our theory satisfies the null
energy condition is sufficient to prove that the collision is always
outside the horizon of the hyperbolic Schwarzschild solution. 
Since
$R$ increases along every timelike path outside the horizon, this
guarantees that the domain wall reaches $R=\infty$.

\doublefig\bous{``Bousso wedges'' indicate in which null directions
the curvature radius of the hyperboloids decreases. Figure (a)
is a piece of AdS spacetime, while (b) is a piece of Minkowski spacetime. The
collision occurs at the point marked with a star($*$). When the
collision occurs at $R<\ell$, it is ``behind the horizon'' in AdS, as
shown in (a).}  
{\epsfxsize2.0in\epsfbox{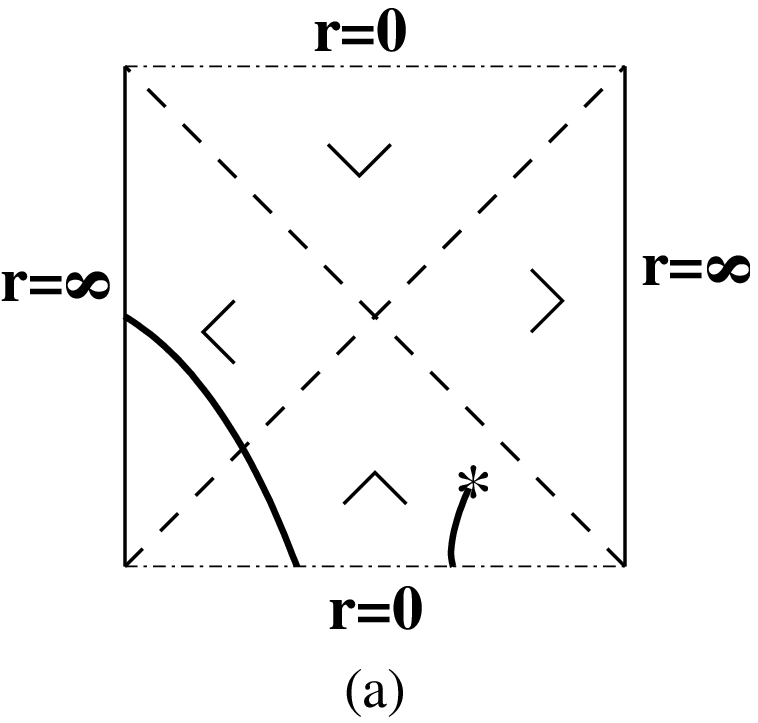}} 
{\epsfxsize2.0in\epsfbox{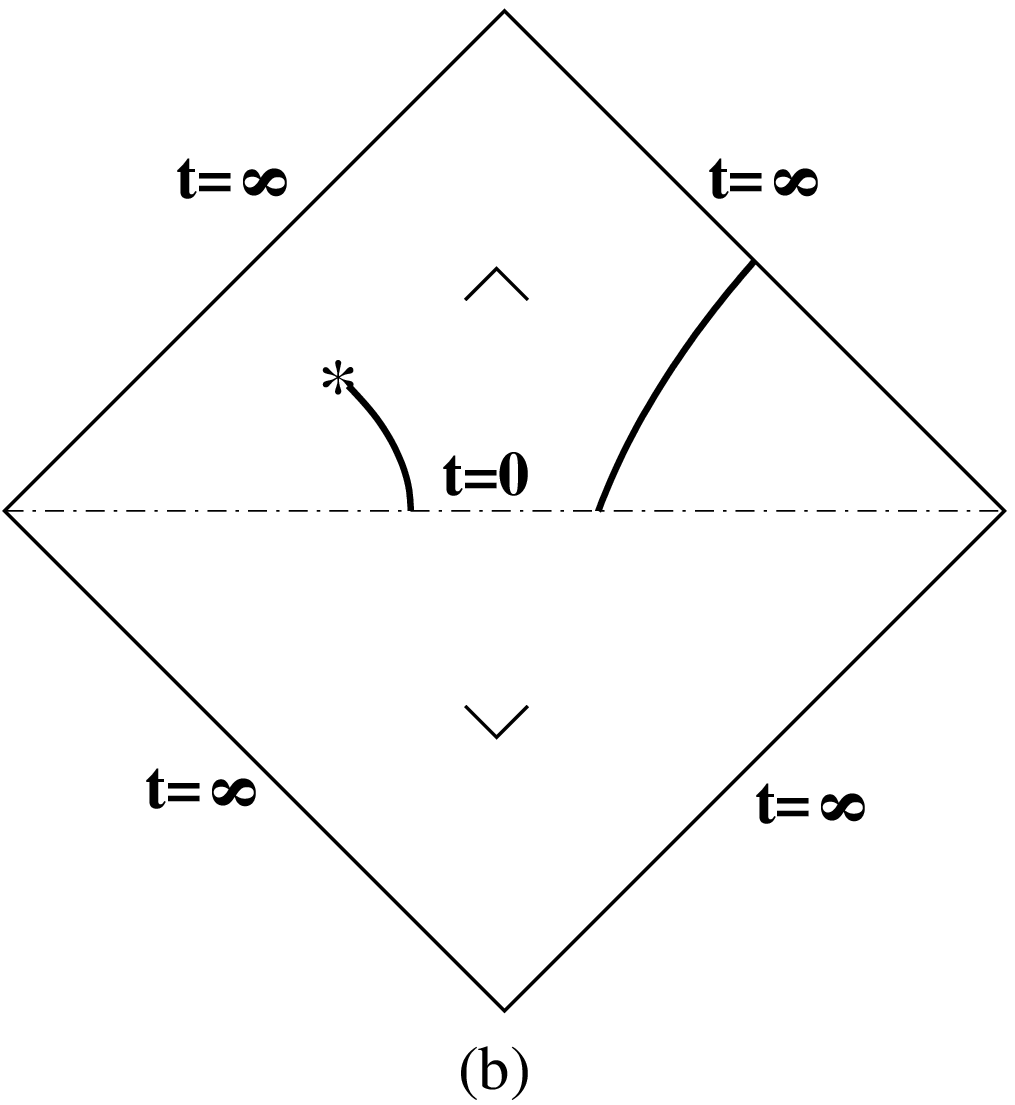}}

To see the
implications of satisfying the null energy condition, it is helpful to
use the notation of ``Bousso wedges''\BoussoXY. Given a spacetime with
$H_2$ symmetry, at any point one can draw four radial null rays. These
are
 null geodesics which are at rest on the $H_2$. Two of these null
rays will be future directed, and two will be past directed. At a
generic point, the radius of curvature of the hyperboloids will
decrease along two of the four rays; along the other two it will
increase. 
To draw the Bousso wedge, we extend lines from the vertex in the
two decreasing directions. For example, if the radius decreases along both
left-moving null geodesics, we draw the wedge ``$>$'' . Recall that
Minkowski space in the $H_2$ slicing has metric
\eqn\minkmetric{
ds^2 = -dt^2 + dz^2 + t^2 dH_2^2}
The radial null geodesics satisfy $dt^2 = dz^2$. For
positive $t$, the size of the $H_2$'s decreases along both
past-directed null rays, while for negative $t$ the the size decreases
along the future-directed rays.
So Minkowski space has the Bousso-Penrose
diagram shown in \bous(b). The diagram for AdS spacetime is shown in
\bous(a).

Now in gluing our spacetimes together, we have two pieces of
information. First, when gluing two spacetimes across a null shell, 
continuity of the
metric 
guarantees that if the null shell is
increasing in radius as seen from one side, it is also increasing
as seen from the other side of the shell. Also, the null energy
condition 
guarantees that if the radius
  is decreasing along a
radial null geodesic, it must continue decreasing until it reaches zero.

\triplefig\wedgefig{This figure demonstrates how to determine the
direction of the Bousso wedge in the spacetime to the future of the
collision, when the collision happens at a radius $R<\ell$. Outside
the future lightcone of the collision, the wedges have the behavior shown in
\bous. Inside the future lightcone, we do not know the direction yet,
as shown in (a). Continuity of $R$ across the null shell partly
determines the behavior inside the future lightcone, as shown in (b). 
Here the dots denote the vertex
of the wedge. Now imagine starting in the $\Lambda=0$ region and
following the known decreasing null ray across the
final domain wall. This ray is left-moving and past-directed.
$R$ must continue to decrease along this ray after crossing the domain wall since we
have assumed the null energy condition, so in the $\Lambda<0$ region
the left-moving past-directed ray must decrease. We can repeat the
same argument reversing the two sides of the domain wall, leading to
the final diagram (c).
}
{\epsfxsize2.0in\epsfbox{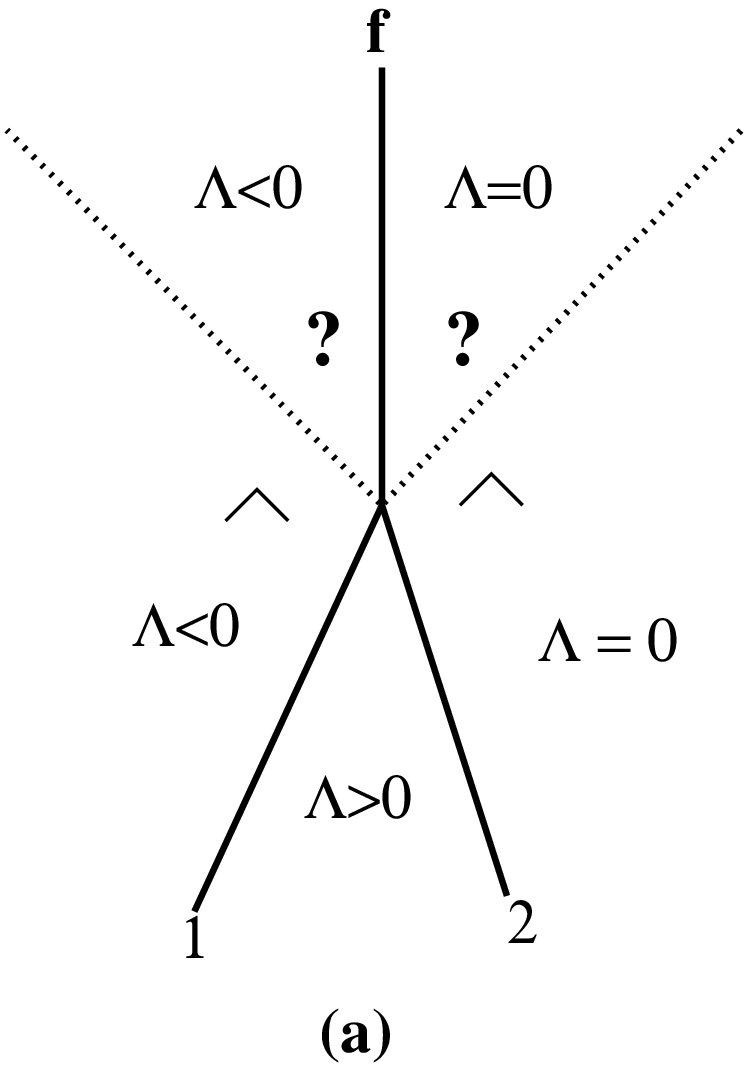}} 
{\epsfxsize2.0in\epsfbox{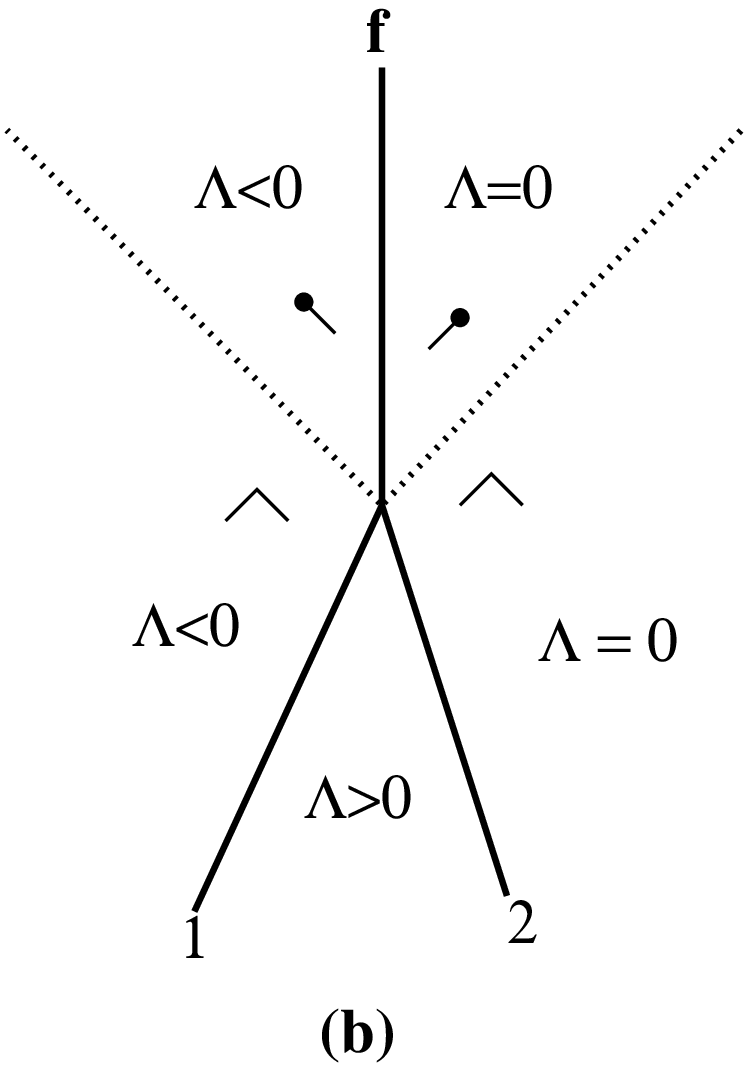}} 
{\epsfxsize2.0in\epsfbox{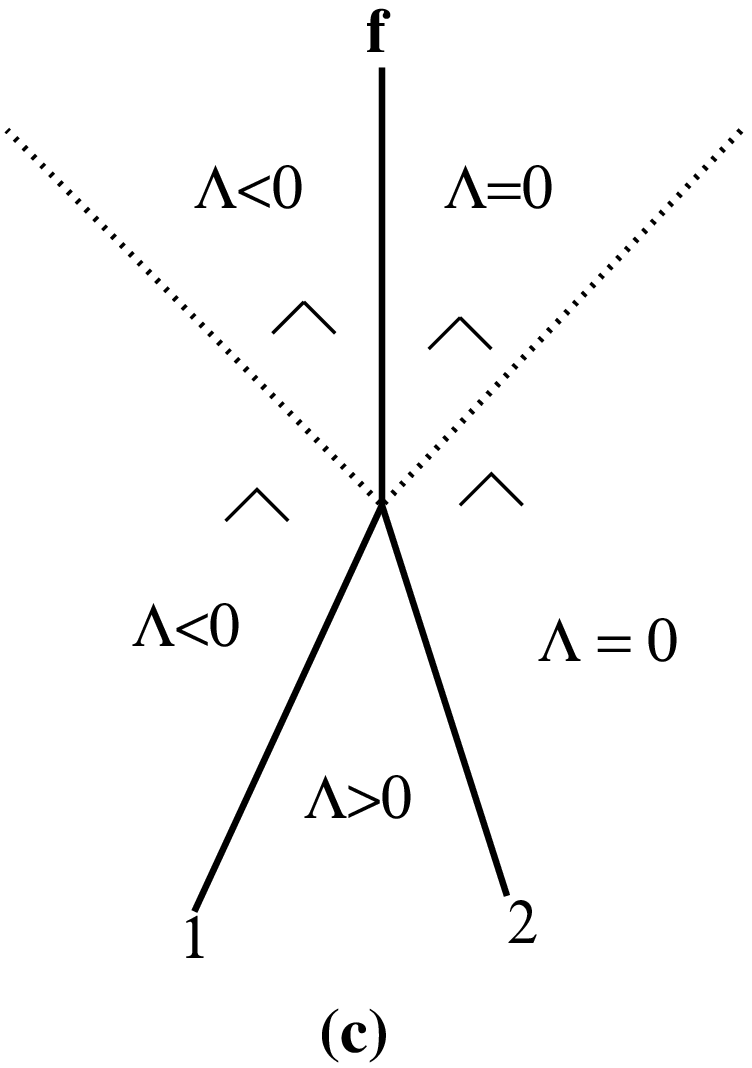}} 

As explained in \wedgefig, when the collision occurs at $R<\ell$ we match
onto a region of the hyperbolic Schwarzschild solution in which $R$ is
decreasing along both past-directed rays, as indicated by the wedge in Fig. 11. 
This is exactly
the region outside the horizon! (The Bousso wedges for the hyperbolic
Schwarzschild solution are shown in \flatbh.) We have also discovered
that $R$ is decreasing along both past-directed rays on the
$\Lambda<0$ side of the domain wall. This information guarantees that
we are inside the ``white hole horizon'', as shown in \smallr{(b). We
have shown at the end of section 2.3 that when the collision occurs
behind the horizon the mass parameter $M$ becomes negative. But as we 
discussed below equation \deff, if $M$ becomes too negative the black hole
horizon disappears altogether, leaving a naked timelike
singularity. Fortunately, such solutions do not contain any region
with the appropriate Bousso wedge $\wedge$, so they cannot be created in the collision.

One can perform a similar analysis when the collision occurs at
$R>\ell$. This case is slightly more subtle. In the AdS spacetime, the collision
now occurs outside the horizon (the region marked "$>$" in \bous(a)). In Minkowski
spacetime nothing has changed so the appropriate wedge is still $\wedge$. 
In the future lightcone of the collision point, the same logic we used above guarantees
that on the $\Lambda<0$ side the wedge is $>$, so the collision now occurs outside the 
horizon of the
black hole. Also, we showed in section 2.3 that in this case the mass parameter M is positive.
On the $\Lambda=0$ side of the domain wall, there are two possibilities which are consistent 
at this level of analysis: $\wedge$ and $>$.  The latter possibility would mean that the collision
occurs behind the horizon of the hyperbolic Schwarzschild solution. However, this possibility
would lead to a relative minus sign in the junction condition, that is, the upper sign in \finalcond.
 However, if the tension
obeys the BPS bound one cannot solve \finalcond\ with a relative minus sign. So $>$ is
inconsistent with the junction condition, and once again the collision must occur outside
the horizon of the hyperbolic Schwarzschild solution in the region $\wedge$.

Since this argument only relies on the null energy condition and the
BPS bound, it is
general. In particular, one could imagine adding excitations to the
domain wall which change its equation of state. Our argument in this
section still shows that the collision occurs outside the horizon of
the hyperbolic Schwarzschild solution. As a result, the domain wall
still must grow to
infinite $R$, where its dynamics will be controlled simply by its
tension relative to the BPS bound, as described in section 2.3.

\newsec{Length Scales}

In this section we discuss the various length scales involved in this problem and what values
they can take in string theory vacua.    In particular we will be interested in justifying the approximations
used in the previous section.  The length scales of interest are:  the curvature length $\ell$
in the AdS vacuum,  $1/\ell^2 \sim \Lambda_{AdS}$; the curvature
length $\ell_{dS}$ in the 
dS vacuum, $1/\ell_{dS}^2 \sim \Lambda_{dS}$; the width of the domain walls $l_w$;  the
initial radius of the nucleated bubbles $R_b$; and the initial separation of the bubbles.

For supersymmetric vacua  we can write 
$\Lambda \sim V/m_p^2 \sim -|W|^2/m_p^4$
where $W$ is the value of the superpotential in the vacuum\foot{For simplicity we suppress the Kahler potential.}
.   In such vacua $\ell \sim m_p^2/W$.
For the thin wall approximation used in the previous section to be valid we need $l_w < \ell, \ell_{dS}$. 

For BPS domain walls 
the tension $\rho$ between vacua 1 and 2 is given  by a relation of the form
\eqn\bpstension{ \rho_{BPS} \sim W_1 - W_2}
 The quantity  $\kappa$ 
discussed in the previous section is related to $\rho$ by $\kappa \sim \rho/m_p^2$.
A BPS domain wall between an AdS and a $\Lambda =0$ supersymmetric vacuuum has tension
$\rho \sim W, ~ \kappa = 1/\ell$.
The domain wall between a de Sitter vacuum and a supersymmetric $\Lambda \le 0$ vacuum
is never BPS, but  the typical examples of  de Sitter vacua in the landscape are ``uplifted" from
some supersymmetric AdS vacuum and the tension of the domain walls need not be much different
from the BPS walls connecting to the AdS vacuum before uplifting.   So we will estimate tensions by
the rough relation $\rho \sim W$.    Further we will make the rough estimate $\Lambda_{dS} \sim
-\Lambda_{AdS}$.  The plethora of landscape constructions indicate that
many other values are possible.

The equations of motion governing the domain wall profile give a relation
${\Delta \phi/ l_w} \sim \sqrt{V}$  where $V$ is the characteristic size of the potential
in the domain wall region and $\Delta \phi$ is the field variation from one side of the domain
wall to the other.  We expect $\Delta \phi \le m_p, ~V \sim \Lambda m_p^2$
so we have a rough upper bound $l_w \le m_p^2/W \sim \ell \sim \ell_{dS}$.   To justify the
thin wall approximation we would like $l_w \ll \ell, \ell_{dS}$.  For vacua localized  down a warped throat the characteristic energy scales are reduced relative to $m_p$, the bulk 4D Planck mass,  by a large warping factor.  We thus expect  moduli excursions between vacua to be small,  $\Delta \phi \ll m_p$, so thin wall  should hold parametrically. Many of the AdS and dS vacua studied in landscape constructions are of this type.  But the  best known $\Lambda = 0$ examples, for instance the large field
vacuum in KKLT \KachruAW\ , are far away in field space from the throat and require $\Delta \phi \sim m_p$.
(In fact they correspond to decompactification limits of Kahler moduli).  Still these domain walls
do not violate thin wall badly, and the results discussed in the previous section should be a reliable qualitative guide.
For parametric control  we can turn to the flux tuned $W=0$ vacua discussed in \DeWolfeNS .
These vacua have moduli but this should not alter our basic picture.  Candidates for isolated
$\Lambda = 0$ are described in  \refs{\BeckerKS,\MicuRD} .

Independent of the string model that produces it, all we need is that
the potential comes from a suitable superpotential, as in \superpot ,
 and that both the scale of
the superpotential and the characteristic field excursion $\Delta
\phi$ are much less than the four-dimensional $m_p$. In this case we
have $l_w \sim (\Delta \phi)^2/W$ and $\ell \sim m_p^2/W$. Putting in
the additional assumption that the de Sitter radius is of order the
$AdS$ radius, we have $l_w \ll \ell, \ell_{dS}$.
It may well be possible to construct a ``string inspired" model of this type using the framework
discussed in \CeresoleIQ .

There is another issue of scales we should discuss.
We have been assuming that the dS bubble decays by Coleman-De Luccia \ColemanAW\
bubble nucleation.   This is the case when $ \rho^2 \ll \Lambda_{dS}~ m_p^4$. 
Our rough estimates are on the edge of this region.
Otherwise the dS vacuum decays
by what is called the Hawking-Moss process \HawkingFZ .   An appealing picture of this process is the quantum 
diffusion of $\phi$ to the top of the barrier and across, then a classical roll 
down \refs{\StarobinskyFX\LindeSK} . One result
of this will be a nonspherical initial bubble on scales bigger than $\ell_{dS}$.  At least for relatively
late collisions where the portion of bubbles involved in the collision is a small fraction of 
the original $\ell_{dS}$ such large scale inhomogeneities should not be important, as discussed below.

A rough estimate of the initial bubble radius $R_b$ is given by the
balance $m_p^2 \Lambda_{dS} R_b^4
\sim \rho R_b^3$.  For the parameter ranges above this gives $R_b \sim \ell_{dS}$.
Coleman-Deluccia is valid when $R_b \leq  \ell_{dS}$.    One way for this to fail is for the 
amount of uplifting to be fine tuned to be small, so $\Lambda_{dS} \ll |\Lambda_{AdS}|$.

 For two
bubbles to collide, the centers can be separated by a maximum geodesic
separation
given by $\pi \ell_{dS}$.
 The de Sitter invariant measure for nucleation
gives the probability distribution for the separation. In the limit
that the nucleation rate is small, one finds a probability density
proportional to \CarvalhoFC
\eqn\prob{ f(r) = \sin\left(r/\ell_{dS}\right) }
where $r$ is the geodesic separation. This means that 
the bubbles are likely to be separated by of order one de Sitter
radius. We have already arranged that the de Sitter radius is large
compared to the thickness of the domain wall.
Since the typical bubble separation is $ \ell_{dS}$, the bubble wall never becomes ultrarelativistic and the scale of energy
liberated in the collision
is set by  $\rho$.  

We can now estimate the size of the parameters $t_0, M$ that appear in the solutions in the previous section using \nullansh,\nullansf, and (A.6).
The domain walls collide on an $H_2$. If the bubbles are separated by
of order one de Sitter radius, then the $H_2$ will have a radius $R \sim \ell_{dS}$.
Under the above assumptions ($\ell \sim  \ell_{dS}$, tensions of order the BPS bound), the parameters $t_0$ and $GM$ turn out to be of order $R$. Thus the collision takes place around the horizon. This is not a problem, but it is comforting to know that there is way to ensure that the collision takes place far outside the horizon. As the bubble separation approaches
its maximum value of $\pi \ell_{dS}$, the radius of curvature of the
$H_2$ at the time of collision becomes very large. Due to the expansion of the de Sitter space between the bubbles,  the energy density at the collision remains bounded, so $\sigma(R)$ approaches a constant. Under these conditions, one can show that the horizon radius satisfies $r_h = a R$ for a constant $a<1$. Thus $R-r_h$ grows with $R$ and the collision is far outside the horizon. This holds for both the BPS and non-BPS case.
On the other hand, if $\ell_{dS} \ll \ell$, then a typical bubble collision will have $R \sim \ell_{dS}$. Since there are no hyperbolic black holes with $r_h \ll \ell$, this collision will take place inside the (past) horizon. As argued in section 2, this leads to $M < 0$. In fact, a typical collision of this type will have $GM \sim - \ell_{dS}$.

\newsec{Possible Complications}
We have seen that within our approximation the Minkowski bubble is not
eaten up by the AdS bubble; specifically, at least part of future null
infinity remains undisturbed all the way up to timelike infinity. We can consider relaxing our
approximation in two ways:

\noindent $\bullet$ What happens when we go beyond the thin wall
approximation?
 
\noindent $\bullet$ What about fluctuations which break the $SO(2,1)$ symmetry?

We argue here that there exists a regime of parameters where these
corrections are unimportant. Furthermore, we believe that even in the
regime where corrections are important our simplified model gives a
correct picture of asymptotic behavior.

The first place we used the thin wall approximation was in assuming that the radiation emitted by the collision comes off in a null shell. However numerical simulations of bubble collisions show that the bubble walls  oscillate for several cycles emitting radiation during each cycle \refs{\HawkingGA, \BlancoPilladoHQ}. (This is suppressed if the scalar field is complex and there are degenerate vacua with a large difference in phase between the scalar field in each bubble.)  One could improve our approximation by having radiation come off in a series of null shells, but this would only cause the parameters $M$ and $t_0$ to increase more slowly. It would not qualitatively change our conclusions.  

We saw in section 3 that by choosing certain landscape vacua, or more concretely
by demanding that our potential come from
a suitable superpotential and asking that the scale of the superpotential be
separated from the four-dimensional Planck scale, we can make the
characteristic thickness of the domain wall small compared to the
curvature scales in the problem. 
As discussed in section 2, the thin wall approximation necessarily
breaks down in the AdS regions of the geometry, leading to crunches
and black holes. The problem is not that the domain wall is thick on
scales of interest, but that the AdS is unstable to infinitesimal
perturbations with the symmetry of our problem. It is the exponential
tails of the domain walls which provide the energy to crunch the AdS space.

There is good reason to believe that even when the domain wall is thick, the evolution will be qualitatively similar to the thin wall results. Consider the asymptotically flat region. The key question is whether a solution with a piece of null infinity necessarily has a complete null infinity (i.e. the null generators can be extended to infinite affine parameter to the future). This is one version of cosmic censorship, and might appear hard to prove. However, the existence of hyperbolic symmetry greatly simplifies the problem. In fact, in a recent paper \DafermosHQ, it was shown that for gravity coupled to collisionless matter, hyperbolic symmetry is enough to prove this form of cosmic censorship. Presumably a similar result holds for  other reasonable matter fields.

Now consider perturbations which break the symmetry of the background.
Such perturbations are inevitable; for example, the domain wall is in
contact with de Sitter space, which produces fluctuations in the shape
of the wall characteristic of the de Sitter temperature. In addition,
when the domain walls collide, there is generically  more energy
than is necessary to form the resulting domain wall. We have considered some of this energy radiating away in a null shell. In addition,  some of it
 can go into excitations of the outgoing domain wall, and these
excitations can be at all wavelengths.

Very long wavelength modes will simply get redshifted away. These can be approximated as a homogeneous perfect fluid and satisfy \conser.  For example, if the energy density is increased by the addition of presureless dust, $ \delta \dot \rho = -2\delta \rho\ \dot R/R$, so $\delta \rho \propto 1/R^2 \rightarrow 0$.
Smaller wavelength modes might coalesce into black holes.
The total energy liberated is of the same order
as that required to make a Schwarzschild black hole of radius $\ell$ in each area of size
$\ell^2$ on the collision surface. 
But as long as the additional asymmetry is small and the collapse time $\ell$ long compared to 
decay times of massive particles the fraction of energy  collapsing will be small.  So the resulting black holes will have  radii much less than $\ell$ and will not affect the dynamics in an important way.

We now briefly consider another possible source of singularities. Even when $\la =0$,
the collision of plane symmetric gravitational waves produces a spacelike singularity everywhere to the future of the collision \KhanVH. As discussed above, when our two bubbles collide, the radius of  curvature  of the hyperboloid of intersection can be very large. Near this surface, the spacetime looks like two almost plane waves colliding. Could this produce a (catastrophic) singularity everywhere to its future? Fortunately, the answer is almost certainly no. First of all, our bubbles are colliding in an expanding de Sitter universe and it has been shown \CentrellaIB\ that in this case,  the collision of plane waves does not generically produce curvature singularities.  More importantly, there is a qualitative difference between between plane symmetry and hyperbolic symmetry (even when the hyperbolae have small curvature): there are no hyperbolic gravitational waves. Without the scalar field, the solutions are the ones discussed in section 2 which do not have spacelike singularities when  $\la =0$. With the scalar field, the dynamics is expected to be similar to the thin wall evolution since the field will radiate away its excess energy.

\newsec{Discussion}

In recent discussions of inflation, bubbles with different cosmological constant can be nucleated. Bubbles with $\la<0$  end in a  crunch, while supersymmetric bubbles with $\la=0$ are expected to last forever. We have studied the effect of a collision with a crunching bubble on a $\la=0$ bubble.  Of course a typical bubble will encounter many collisions with other bubbles. If a  $\la<0$ bubble is entirely surrounded by $\la=0$ bubbles, the crunching region is enclosed inside a two-sphere and probably forms an ordinary spherical black hole, which can then evaporate. Conversely, if a $\la=0$ bubble is entirely surrounded by crunching bubbles, then there is no asymptotically flat region. Since the energy density inside the bubble is greater than outside and the surface tension is positive, the bubble will probably collapse and form a big crunch everywhere.

Generically, a bubble of one type will not be entirely surrounded by bubbles of the other type. To summarize the effect of multiple collisions  consider future null infinity of the $\la=0$ bubble.  Before collisions this has the same structure as the future null infinity of Minkowski space, a null line cross a spatial sphere. If the tension in the final domain wall is above the BPS bound  ($\kappa > 1/\ell$)  it is clear from Fig. 5 that  a part of null infinity including the  full  asymptotic spatial sphere is left
undisturbed.  In the BPS case ($\kappa = 1/\ell$), on the other hand,  it is clear from Fig. 8 that just one collision removes part of the asymptotic spatial sphere.  Later collisions remove successively smaller portions of the asymptotic sphere, eventually
producing a fractal.  Only a set of measure zero is left undisturbed.
Such fractals are standard in eternal inflation; a similar sequence of collisions was discussed in \BoussoGE.
 This structure is relevant for the Conformal Field Theory picture of eternal inflation discussed in \FreivogelXU .  It would be interesting to know how often such BPS domain walls occur in the landscape.

We can also ask what  a timelike observer in the $\la = 0$ FRW bubble sees at late times on his sky.
This is relevant for the ``census taker'' approach to eternal inflation measures discussed in \censustaker .
For both $\kappa > 1/\ell$ and $\kappa = 1/\ell$ this observer will see a fractal of black disks on the sky.
 The $\kappa > 1/\ell$
domain walls will be accelerating away from the observer at a constant rate, while the $\kappa =1/\ell$
walls will only be accelerated by the cosmological expansion, a rate which  decreases with time.  The 
acceleration radiation from the $\kappa > 1/\ell$ walls should provide a distinctive signature of this behavior.

\vskip 1cm
\centerline{\bf Acknowledgements}
\vskip .5cm
We would like to thank Raphael Bousso, Shamit Kachru, Renata Kallosh, Alex Maloney, Rob Myers,
Eva Silverstein, and Lenny Susskind for valuable discussions.
This work was supported in part by The Stanford Institute for Theoretical Physics, the
Berkeley Center for Theoretical Physics, DOE grant DE-AC03-76SF00098, and NSF grants PHY-0555669, 9870115, 0455649.

\appendix{A}{Conservation of energy at collision}

At the collision point in the thin wall approximation, we must impose energy and momentum
conservation. Since we are constructing solutions of general
relativity, energy-momentum conservation will be automatic as long as
we are careful to solve Einstein's equation even at the collision
point. 
It turns out that there
is only one condition which needs to be imposed: the geometry should
not have a conical singularity at the collision point. Because 
we are using different metrics in different patches around the
collision point, imposing this condition is not trivial. \LangloisUQ\
described a convenient method for writing the condition, and we are
able to further simplify their method.

Given a number of domain walls colliding at a point, one can compute
the boost angle $\beta$ between two adjacent walls. More precisely,
choose points on the domain walls which are at rest on the $H_2$,
and compute their relative boost. This is
straightforward between two incoming walls, or between two
outgoing walls; for example, one can define the boost angle by
\eqn\boost{ u_1 \cdot  u_2 = |u_1| |u_2|  \cosh \beta_{12}}
where $u_i$ are the 4-velocities.
Between an incoming and an outgoing wall, we define the boost angle to
be negative,
\eqn\boostneg{ u \cdot v = |u| |v| \cosh(-\beta)}
Note that each of these boost angles can be computed within a single
coordinate system, because we only need to compute the boost between
adjacent domain walls. The condition of regularity at the collision point
is simply that the boost angles should sum to zero:
\eqn\reg{ \sum_{i = 1}^n \beta_{i,i+1} = 0}

\ifig\conical{The absence of a conical deficit at the collision point leads to a constraint on the relative boosts $\beta$ of the colliding domain walls, equation \reg. }
{\epsfxsize2.5in\epsfbox{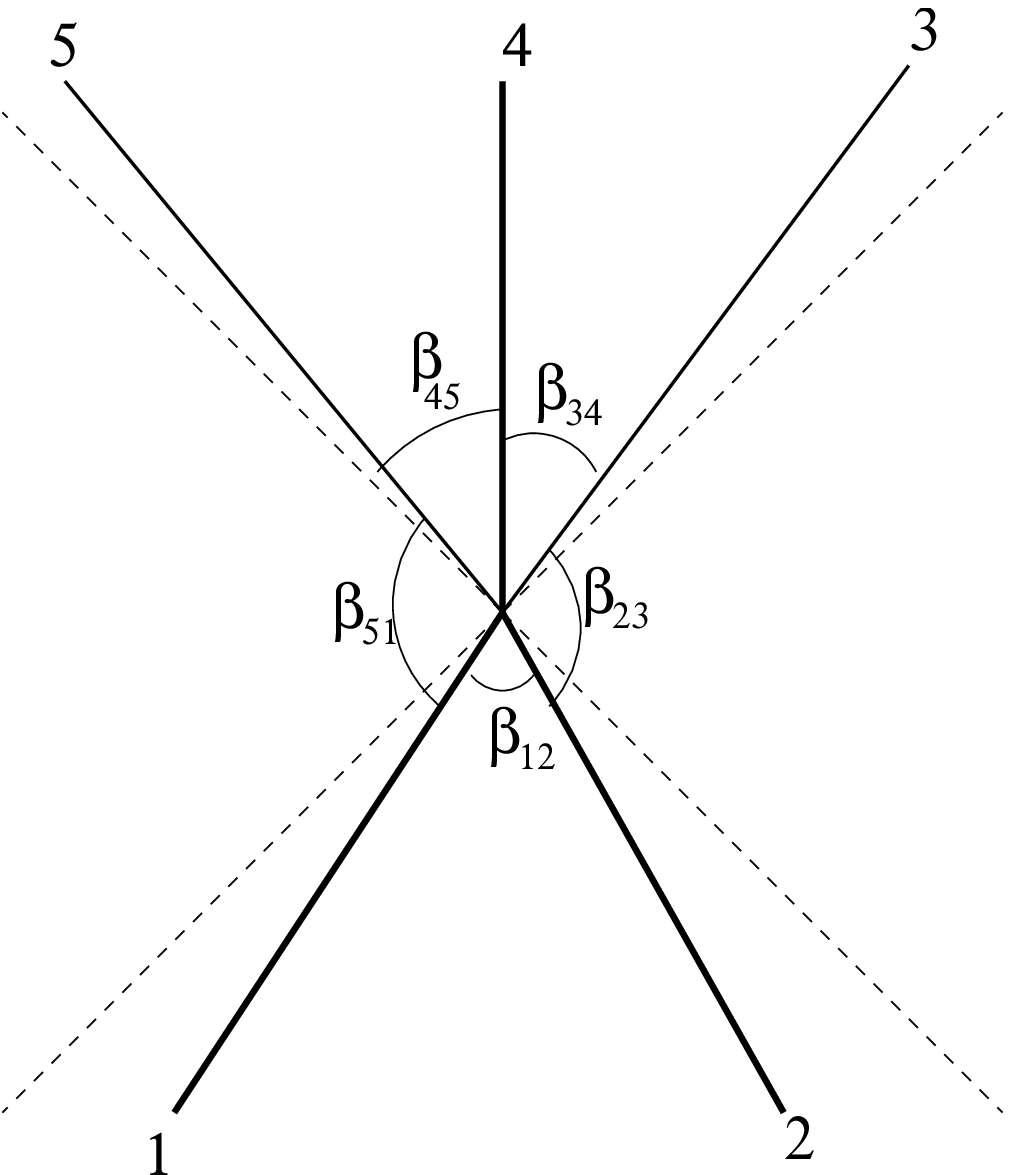}}
An example is illustrated in \conical .
One can include null shells in this formalism by taking an appropriate
limit of a timelike domain wall; in this limit some of the boost
angles will diverge but the sum of the boost angles will remain
well-defined. 

In the case of interest, we have two incoming domain walls and one
outgoing domain wall, and two outgoing null shells of radiation. Each
domain wall is characterized by the proper time derivative of the size
of the $H_2$ along the domain wall, which we denote by $\dot R_1, \dot
R_2$ for the incoming walls and $\dot R_f$ for the outgoing wall. The
regularity condition can be written in terms of these ``velocities''
and the metrics of spacetimes which are glued together. After some
algebra, we find
\eqn\regeq{\eqalign{
\cosh^{-1} {\dot R_1 \over \sqrt{g}}  - \sinh^{-1} {\dot R_1 \over \sqrt{f_0}}
+  \cosh^{-1} {\dot R_2 \over \sqrt{g}}  - \cosh^{-1} {\dot R_2} 
+ \cosh^{-1} {\dot R_f \over \sqrt{h}}
 + \sinh^{-1} {\dot R_f  \over \sqrt{f}} = {1 \over 2} \log{h f \over f_0}
}}
with the definitions
\eqn\alphadef{ \eqalign{
g &= 1 + {R^2 \over \ell_{dS}^2} \ \ \ \ \ \ \ \ \ \ \ \ \ \ \ \ \ \ \
\ \ \ \ \ \ 
h = 1 - {t_0 \over R} \cr 
f &= -1 + {R^2 \over \ell^2} - {2 G M \over R} \ \ \ \ \ \ \ \ \
\ 
f_0 = -1 + {R^2 \over \ell^2} \cr
}}
Note that, since $\cosh^{-1}$ and $\sinh^{-1}$ are both simply related to logarithms, 
\regeq\ can be rewritten as an algebraic equation.

At this point, it is confusing how many free parameters exist in the
solution. We can rewrite \regeq\ in a way which clarifies this:
\eqn\regclar{
\cosh^{-1} {\dot R_f \kappa R \over \sqrt {h f}} + {1 \over 2}
\log {f_0 \over h f} = \cosh^{-1} {\dot R_1 \kappa_1 R \over
\sqrt {f_0  g}} + 
\sinh^{-1} {\dot R_2 \kappa_2 R \over \sqrt g}
}
Here $\kappa$ is related to the tension of the final domain wall as
defined earlier,
$\kappa = 4 \pi G \rho$; $\kappa_1$ and $\kappa_2$ are defined similarly.

A good way to think about the number of free parameters is the following. First, take
all of the parameters of the underlying theory as fixed: the tensions
of all domain walls $\kappa_i$, and the cosmological constants. Also, fix one initial
condition, the separation between the bubbles. Given this data, the
radius of curvature at the collision $R$ and the ``velocities'' at
the collision $\dot R_1, \dot R_2, \dot R_f$, are determined by the Israel
junction conditions. We now have two unknown parameters, which
physically correspond to the energy density in each of the shells of
radiation. Equivalently, the two unkowns are the two ``mass'' parameters
 $t_0$ and $M$, which are hiding in the functions $h$ and $f$ in \regclar. 
The regularity condition \regclar\ provides one constraint
on these two parameters. 
Microphysics, e.g., the form of the scalar potential, clearly fixes all these
parameters.  In the absence of a full solution of the field equations it seems reasonable
to assume that the shells of radiation have roughly equal energy densities.

\listrefs

\end